%% file: main.tex
\documentclass[%
aps,
 superscriptaddress,
 twocolumn,
 showpacs,
preprintnumbers,
nofootinbib,
 amsmath,amssymb,
 prd,
balancelastpage,
nofootinbib,
floatfix
]{revtex4-1}
\usepackage[utf8]{inputenc}

\usepackage{graphicx}
\usepackage{dcolumn}
\usepackage{bm}
\usepackage{amsmath,amssymb}
\usepackage{diagbox}
\usepackage{rotating}
\usepackage{multirow}
\usepackage{xcolor}
\usepackage{hepunits}
\usepackage{comment}
\usepackage{hyperref}
\usepackage{enumitem}

\newcommand{\exclude}[1]{}

\newcommand{\beq}{\begin{equation}}
\newcommand{\eeq}{\end{equation}}
\newcommand{\ba}{\begin{aligned}}
\newcommand{\ea}{\end{aligned}}
\newcommand{\bpm}{\begin{pmatrix}}
\newcommand{\epm}{\end{pmatrix}}
\newcommand{\vpi}{\vec{p}_i}
\newcommand{\vpf}{\vec{p}_f}

\begin{document}

\preprint{FERMILAB-PUB-22-705-PPD-QIS-T}

\title{Measuring the Migdal effect in semiconductors for dark matter detection}

\author{Duncan~Adams}\thanks{duncan.adams@stonybrook.edu}
\affiliation{C.N. Yang Institute for Theoretical Physics, Stony Brook University, NY 11794, USA}

\author{Daniel~Baxter}\thanks{dbaxter9@fnal.gov}
\affiliation{Fermi National Accelerator Laboratory, Batavia, IL 60510, USA}

\author{Hannah~Day}\thanks{hjday2@illinois.edu}
\affiliation{Department of Physics, University of Illinois Urbana-Champaign, Urbana, IL 61801, USA}

\author{Rouven~Essig}\thanks{rouven.essig@stonybrook.edu}
\affiliation{C.N. Yang Institute for Theoretical Physics, Stony Brook University, NY 11794, USA}

\author{Yonatan~Kahn}\thanks{yfkahn@illinois.edu}
\affiliation{Department of Physics, University of Illinois Urbana-Champaign, Urbana, IL 61801, USA}
\affiliation{Illinois Center for Advanced Studies of the Universe, University of Illinois Urbana-Champaign, Urbana, IL 61801, USA}

\date{\today}

\begin{abstract}

The Migdal effect has received much attention from the dark matter direct detection community, in particular due to its power in setting leading limits on sub-GeV particle dark matter. 
However, it is crucial to obtain experimental confirmation of the Migdal effect through nuclear scattering using Standard Model probes. 
In this work, we extend existing calculations of the Migdal effect to the case of neutron-nucleus scattering, with a particular focus on neutron scattering angle distributions in silicon. 
We identify kinematic regimes wherein the assumptions present in current calculations of the Migdal effect hold for neutron scattering, and demonstrate that these include viable neutron calibration schemes. 
We then apply this framework to propose an experimental strategy to measure the Migdal effect in cryogenic silicon detectors using an upgrade to the NEXUS facility at Fermilab.

\end{abstract}

\maketitle


A proliferation of direct detection experiments searching for sub-GeV dark matter (DM) has been matched by a suite of theoretical work to better understand the kinematics of low-energy scattering in the regime where particle physics and condensed matter intersect~\cite{Kahn:2021ttr}.
This kinematic regime primarily differs from traditional WIMP scattering in that the energy and momentum transfers involved are comparable to the fundamental scales of the target (set by the gap energy and inverse atomic size, respectively), meaning that standard elastic scattering approximations~\cite{Lewin:1995rx} no longer hold.
Indeed, the primary scattering channel of interest for sub-GeV DM searches has long been DM-electron scattering~\cite{Essig:2011nj}, which must account for both the inherent binding energy of the scattered electron and the band structure of the target.
More recently, several theoretical advancements have uncovered yet another inelastic scattering channel of interest for sub-GeV DM, nuclear recoils that directly ionize the scattered atom, a process denoted the ``Migdal effect" (ME).

The theoretical underpinnings of the ME go back to the early work of Arkady Migdal~\cite{Migdal1939,Migdal1941}, who calculated the probability that a radioactive decay would directly ionize the daughter nucleus. Such ionization has been measured in radioactive decay, and is more commonly referred to as ``electron shake-off"~\cite{Carlson,Rapaport,Couratin}. 
Though a handful of papers~\cite{Vergados:2004bm,Moustakidis:2005gx,Bernabei:2007jz} pointed out the likely relevance of this effect for DM-nucleus scattering, progress on the ME accelerated after Ref.~\cite{Ibe:2017yqa} derived the necessary electronic excitation probabilities relevant for DM experiments. A flurry of theoretical activity~\cite{Dolan:2017xbu,Bell:2019egg,Baxter:2019pnz,Essig:2019xkx,Liang:2019nnx,Liu:2020pat,Kahn:2020fef,Knapen:2020aky,Liang:2020ryg,Flambaum:2020xxo,Bell:2021zkr,Acevedo:2021kly,Wang:2021oha,Liang:2022xbu,Blanco:2022pkt} followed, expanding the theory of the ME for galactic DM scattering in isolated atom~\cite{Liu:2020pat}, molecular~\cite{Blanco:2022pkt}, and solid-state~\cite{Liang:2020ryg,Knapen:2020aky,Liang:2022xbu} targets, as well as for solar coherent elastic neutrino-nucleus scattering~\cite{Bell:2019egg}. 
The ME was also shown to dominate over another important inelastic channel, namely the bremsstrahlung process~\cite{Kouvaris:2016afs,Bell:2019egg}. 
Several experimental collaborations have since used these theoretical results to set what are currently the strongest limits on DM-nuclear scattering below $\sim$1~GeV~\cite{LUX:2018akb,EDELWEISS:2019vjv,CDEX:2019hzn,XENON:2019zpr,COSINE-100:2021poy,SuperCDMS:2022kgp,DarkSide:2022dhx}.
Out of all of this work from the DM community, only Refs.~\cite{Nakamura:2020kex,Liao:2021yog,Bell:2021ihi,Araujo:2022wjh,Cox:2022ekg} have so far explicitly considered the ME for neutron scattering. 
A parallel effort in chemistry focused on neutron scattering in isolated atoms and molecules~\cite{lovesey1982electron,elliott1984study}, in part to explain anomalously large neutron cross sections on hydrides~\cite{gidopoulos2005breakdown,reiter2005origin,colognesi2005can}.
Our work differs from these proposals by focusing on the angular distribution of neutrons scattered from solid-state targets.\footnote{See, however, Ref.~\cite{johnson1982inelastic} which employed a similar setup to perform the first inelastic scattering measurements of eV-energy neutrons from liquid targets and which motivated the first derivation of the ME in molecules~\cite{lovesey1982electron}, though the neutron energies were too low to observe the Migdal signal.}
While the existence of the ME is well founded, the magnitude of the effect must be measured to understand the expected DM signal in a direct detection experiment. 

In this Letter, we highlight many of the subtle differences in the ME between the cases of sub-GeV DM and traditional neutron probes. 
These differences arise because sub-GeV DM is lighter than the neutron and thus carries less momentum than a neutron of the same kinetic energy. 
We carefully delineate the theoretical approximations made when calculating ME rates to define the regime where they continue to hold for neutron scattering, which differs considerably from the regime of validity for DM scattering depending on the neutron energy.  
We expand the framework of Ref.~\cite{Bell:2021ihi} to include the angular dependence of neutron scattering, as is used in standard neutron calibration experiments involving neutron detector backing arrays. 
A key finding of this study is that, for an isotropic target in the limit of small momentum transferred to the electronic system (the ``soft limit''), the angular distribution of the scattered neutron factorizes from the electronic matrix element, allowing for a direct calibration of this matrix element for DM scattering. 
We demonstrate that the electronic matrix element for the ME can be measured with a greatly reduced or even completely absent elastic scattering background by a judicious choice of the neutron beam energy and the neutron scattering angle.
However, the expected rates we find for such a measurement are at the boundary of what is currently feasible with existing setups and techniques. 
We conclude that, in order to measure the ME with neutrons in semiconductors, a dedicated low-energy neutron calibration setup is required, and propose one such experiment using modifications to the existing NEXUS facility~\cite{Battaglieri:2017aum} at Fermi National Accelerator Laboratory (Fermilab).


The ME is defined as the ionization or excitation of an atomic electron accompanying the recoil of the atom's nucleus~\cite{Migdal1941}. 
For sub-GeV DM, the ME greatly enhances the sensitivity of direct detection experiments to the DM-nucleon cross section~\cite{Baxter:2019pnz,Essig:2019xkx,Essig:2022dfa} because the electronic excitations are observable even when the nuclear recoil is below threshold.
For both isolated atoms and semiconductors, under various sets of assumptions (which will be discussed further below and delineated in detail in Appendices~\ref{app:softlimit} and ~\ref{app:nosoft}), the ME rate spectrum $R_M$ factorizes into a quasi-elastic nuclear recoil rate $R_{el}$ and an electronic excitation probability $d\widetilde{P}_e/d\omega$, such that
\begin{equation}
\label{eq:RMGeneral}
\frac{d^2 R_M}{dE_r d\omega} = \frac{dR_{el}}{dE_r} \times q^2 \frac{d\widetilde{P}_e}{d\omega}.
\end{equation}
Here, $E_r$ is the nuclear recoil energy, $\omega$ is the total energy deposited in the electronic system (excitation or ionization), and $\vec{q} = q \hat{q}$ is the momentum transfer from the neutron probe to the target. 
For both classes of targets, the electronic spectrum scales as $q^2$, and we have explicitly factored out this scaling. The goal of this Letter is to devise a scheme to measure $d\widetilde{P}_e/d\omega$ in semiconductors.

For isolated atoms, the electronic ionization spectrum is~\cite{Ibe:2017yqa} 
\begin{equation}
\label{eq:ibe}
     \left ( \frac{d\widetilde{P}_e}{d\omega} \right )_{\rm atom} = \left(\frac{m_e}{m_N}\right)^2 \frac{1}{2\pi}\sum_{i,f} |\langle \psi_f^{(\omega)} | \hat{q} \cdot \vec{r}_e |  \psi_i \rangle|^2,
\end{equation}
where $m_e$ and $m_N$ are the electron and nucleus mass, respectively, $\vec{r}_e$ is the electron position operator, the sum runs over initial and final single-electron orbital quantum numbers, and the final state is a spherical wave with wavenumber $k = \sqrt{2m_e (\omega - |E_b|)}$ where $E_b$ is the binding energy of the initial state. Eq.~(\ref{eq:ibe}) was derived within the context of the Born-Oppenheimer approximation~\cite{Ibe:2017yqa}, but in fact does not require this assumption and is correct to $\mathcal{O}(m_e/m_N)^2$~\cite{Kahn:2021ttr,lovesey1982electron}. 
On the other hand, Eq.~(\ref{eq:ibe}) \emph{does} assume $q \ll m_N/(m_e a_0) \simeq 200 \ {\rm MeV} \left(\frac{m_N}{26 \ {\rm GeV}}\right)$ where $a_0$ is the Bohr radius, since it was derived from the dipole approximation to the exponential $\exp \left(i\frac{m_e}{m_N} \vec{q} \cdot \vec{r}_e\right)$. 
Note that the dependence on $\hat{q}$ in Eq.~(\ref{eq:ibe}) drops out when summing over spherically-symmetric filled electron shells, such that $d\widetilde{P}_e/d\omega$ is isotropic.

In a solid-state system, the ME derivation must be modified because the constituent atoms are no longer free, which gives a characteristic energy scale $\omega_{ph}$ for optical phonon excitation ($\approx 10--100 \ {\rm meV}$ in typical solid-state systems~\cite{Kahn:2021ttr}) and also removes the constraint of exact momentum conservation for the nucleus because the atoms are no longer in momentum eigenstates. 
However, the form of Eq.~(\ref{eq:RMGeneral}) can be recovered under the following three assumptions:
\begin{itemize}
\item Impulse approximation: $q \gtrsim \sqrt{2 m_N \omega_{ph}}$
\item Free-ion approximation: initial nucleus state is a zero-momentum plane wave
\item Soft limit: $k \ll q$ and $\vec{q} \cdot \vec{k} \ll m_N \omega$, where $\vec{k}$ is the momentum transferred to the electronic system.
\end{itemize}
The result for a solid-state system is~\cite{Knapen:2020aky}
\begin{equation}
\label{eq:Lin}
\left (\frac{d\widetilde{P}_e}{d\omega} \right )_{\rm sol.} = \frac{4 \alpha }{\omega^4 m_N^2} \int \frac{d^3 \vec{k}}{(2\pi)^3} Z_{\rm ion} ^2(k)(\hat{q} \cdot \hat{k})^2 \,\mathcal{W}(\vec{k}, \omega),
\end{equation}
where $Z_{\rm ion}(k)$ is an effective momentum-dependent charge of the nucleus plus inner-shell electrons, $\alpha \simeq 1/137$ is the fine structure constant, and $\mathcal{W}(\vec{k}, \omega)$ is the energy loss function (ELF) of the target which measures its response to charge perturbations. 
If the ELF is isotropic and only depends on the magnitude of $\vec{k}$, the dependence on $\hat{q}$ drops out, as in the atomic case. We have verified that for all of the kinematic configurations we will consider, the impulse approximation is valid.

\begin{figure*}
    \centering
    \includegraphics[width=1.70\columnwidth]{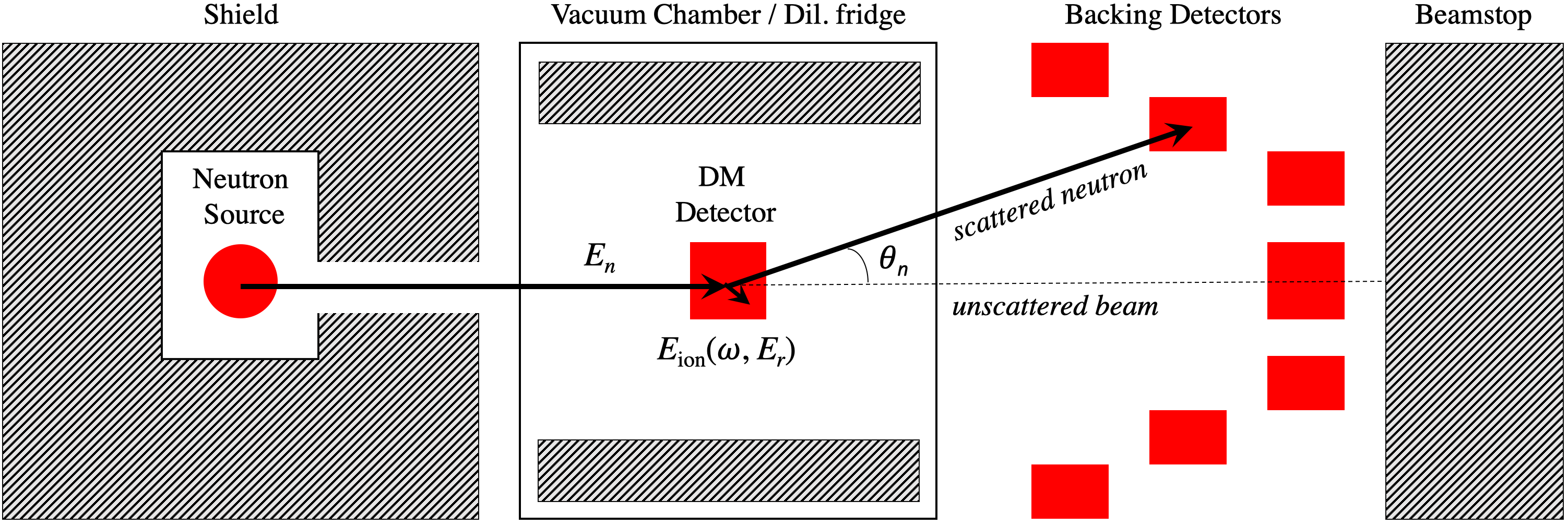}
    \caption{A diagram of an ideal neutron scattering experiment with a backing array, which consists of a series of active (red) and passive (grayscale) elements. Neutrons are generated isotropically from a source placed inside of a shield with a small opening to collimate the beam. These neutrons then enter the vacuum chamber (often a dilution refrigerator) with energy $E_n$, and scatter with lab-frame angle $\theta_n$ into a circular backing array element after transferring $E_r$ of energy to the nuclear recoil and energy $\omega$ to electrons, which together are detected as ionization energy $E_{ion}$. Unscattered neutrons, meanwhile, pass through a capture detector (e.g.~$^3$He counter) to help normalize the simulated beam flux before arresting in a beamstop.}
    \label{fig:diagram}
\end{figure*}

To demonstrate the main kinematic features of the ME, we model the valence shell of silicon with Eq.~(\ref{eq:Lin}), using the isotropic \texttt{GPAW} ELF from \texttt{DarkELF}~\cite{Knapen:2021bwg}, which has a regime of validity $\omega \lesssim 75 \ {\rm eV}$, $k \lesssim 22 \ {\rm keV}$. For larger $\omega$, we model the inner-shell electrons with Eq.~(\ref{eq:ibe})~\cite{Ibe:2017yqa}. 
This division is somewhat artificial, and we discuss its limitations in Appendix~\ref{app:nosoft}. 
As we show in Appendices~\ref{app:chargevar} and~\ref{app:softlimit}, in the soft limit, we can convert the nuclear recoil spectrum into an angular spectrum: 
\begin{equation}\label{eq:result}
    \frac{d^2P_M}{d\cos\theta_n d\omega} = 
    \frac{d\widetilde{P}}{d\cos\theta_n} \big( E_n, \omega, \cos\theta_n \big)  \frac{d \widetilde{P}_e}{d\omega}\big( \omega \big),
\end{equation}
where $E_n$ is the kinetic energy of the incident neutron, $\theta_n$ is the lab-frame angle of the scattered neutron, $\tilde{P}$ is a kinematic prefactor containing all angular dependence, and we have expressed the spectrum as a differential probability $P_M$ of Migdal scattering per incident neutron (rather than a flux-dependent rate). 
Fig.~\ref{fig:diagram} illustrates these kinematics and the experimental setup. 

In Eq.~(\ref{eq:result}), we have explicitly noted the separation of the ionization (ME) probability and the kinematic prefactor containing the angular dependence, and absorbed the $q^2$ scaling from Eq.~\eqref{eq:RMGeneral} into $\widetilde{P}$ such that it now has units of (events/neutron)$\times$[eV]$^2$. 
This unconventional choice of normalization allows us to group all the terms depending on the experimentally-controllable variables $E_n$ and $\cos \theta_n$ together in the explicit expression 
\begin{equation}\label{eq:lab}
\begin{split}
     \frac{d\widetilde{P}}{d\cos\theta_n}  &= \frac{N_0 \rho_T L \sigma_{el}}{A_N}  \frac{ \mu^2 m_N E_n}{\beta m_n^2} \left( \frac{m_n}{m_N} \cos\theta_n  + \beta \right)^2   \\
    & \times \left\{ 1- \frac{\mu^2}{m_n^2} \left( \frac{m_n}{m_N} \cos\theta_n + \beta \right)^2 - \frac{\omega}{E_n} \right\}
    , 
\end{split}
\end{equation}
where $m_n$ is the neutron mass, $\mu \equiv m_n m_N/(m_n + m_N)$ is the reduced mass, $\sigma_{el}$ is the elastic neutron cross section on a target material with density $\rho_T$ and thickness $L$ in the beam direction, $A_N$ is the target's atomic mass number, $N_0$ is Avogadro's number, and
\begin{equation}
    \beta \equiv \sqrt{1 - \frac{ m_n^2}{m_N^2}(1 - \cos^2\theta_n)-\frac{m_n \omega}{\mu E_n}}.
    \label{eq:betadef}
\end{equation}

In Appendix~\ref{app:nosoft}, we demonstrate that the factorization of Eq.~\eqref{eq:result} that allows this separation does \emph{not} hold outside of the soft limit for a semiconductor. We therefore note that calibrating the semiconductor ME outside of the soft limit, for example with high-energy (MeV-scale) neutrons, is fundamentally no longer probing the same regime as sub-GeV DM-nucleus scattering, where the soft limit approximations always hold. 
For the proposed calibrations discussed in the rest of this Letter, we will fall safely within the soft limit (see Appendix~\ref{app:softlimit}), and thus the results of any such calibration are effectively a measurement of $d\widetilde{P}_e/d\omega$ that can be directly translated to DM-nucleus scattering.


Since we do not measure $\omega$ directly, we change variables again to the observable $E_{ion}$, the total amount of energy available as ionization, defined as
\begin{equation} \label{eq:quenching}
    E_{ion} \equiv \omega + f_n(E_{r}) E_{r},
\end{equation}
where $f_n(E_r)$ denotes the ionization efficiency for elastic nuclear recoils as a function of the nuclear recoil energy $E_r(\omega,\theta_n)$. 
For the purposes of this work we consider the Sarkis model~\cite{sarkis2022} as a best theoretical approximation for the ionization efficiency in the mostly uncalibrated regime of small $E_{r}$~\cite{Chavarria:2016xsi}. In general, calibrations of the ME will be dependent on this ionization efficiency (quenching) model; however, we propose two specific calibration schemes in this Letter designed to minimize this dependence. A more complete treatment would also consider the systematic or theoretical fluctuations in $f_n(E_r)$, but this is outside the scope of this work. 

\begin{figure*}
    \centering
    \includegraphics[width=0.49\textwidth]{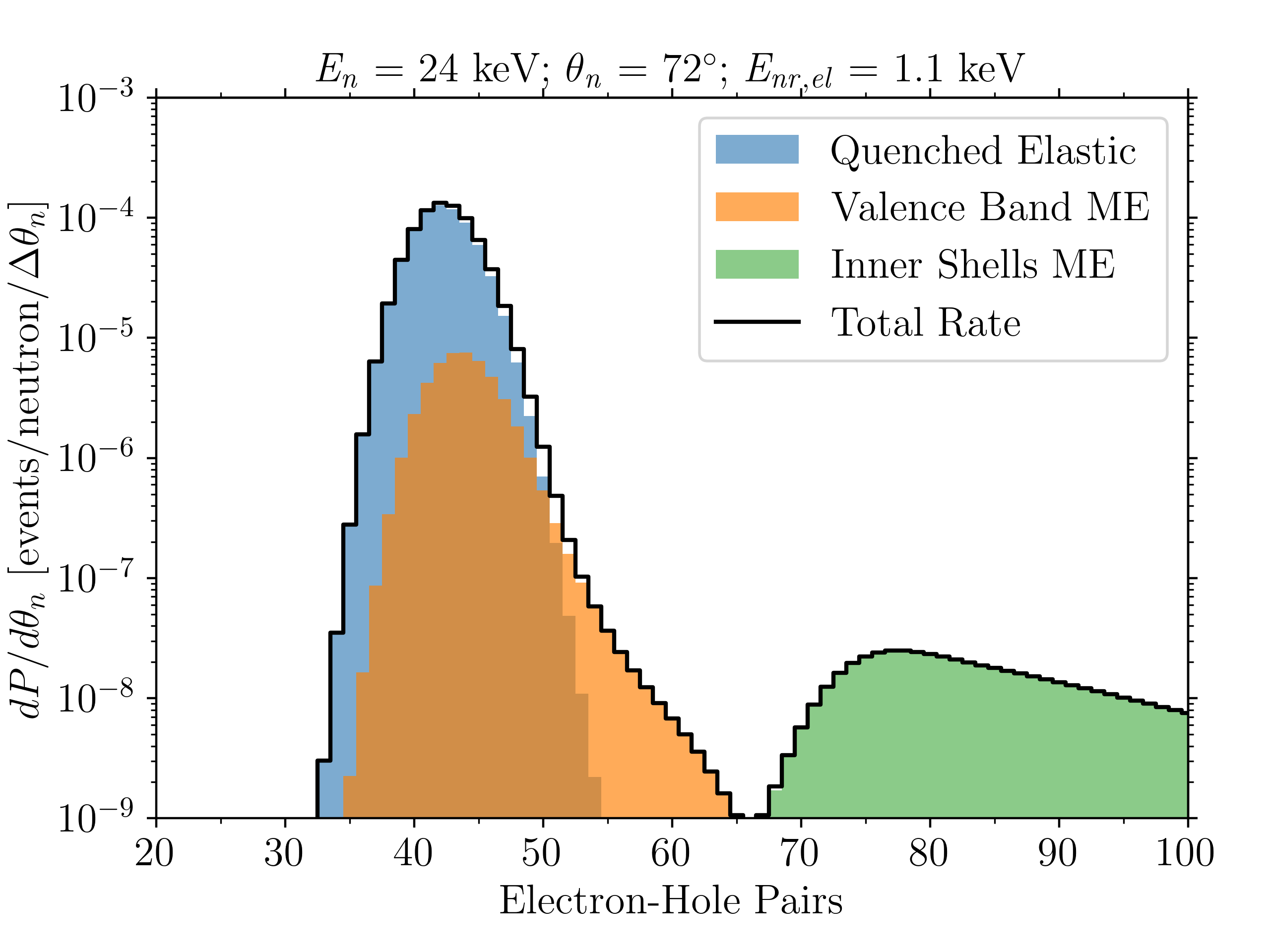}~\includegraphics[width=0.49\textwidth]{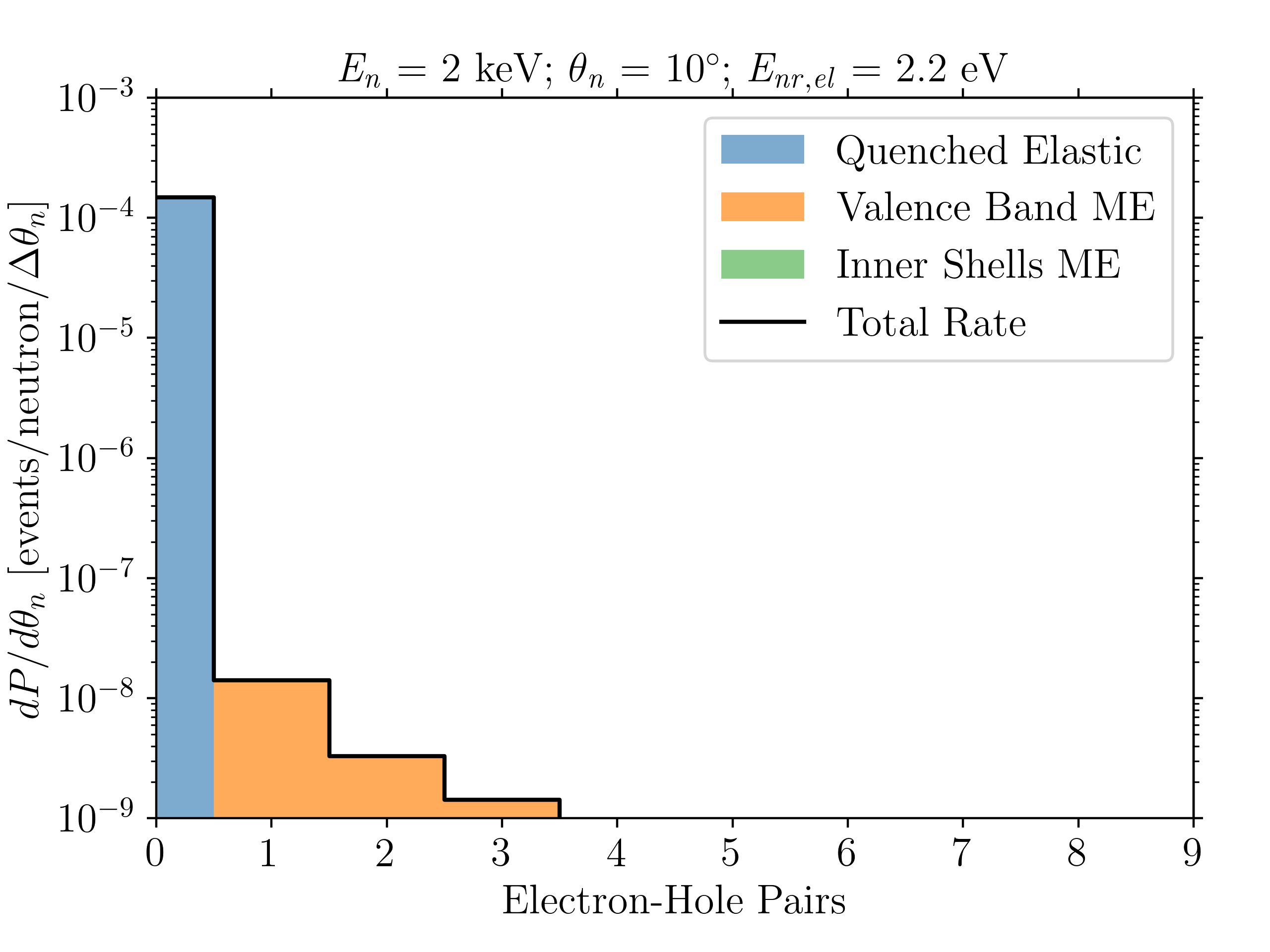}
    \caption{ 
    Differential probability spectra $dP_{n_e}/d\theta_n$ (in units of events/neutron/degree of angular coverage) are shown per detectable charge quanta $n_e$ in the left~(right) plot for an ideal 1~cm thick silicon detector in a $E_n = 24$~(2)~keV monoenergetic neutron beam at a fixed scattering angle of $\theta_n = 72$~(10)~degrees, assuming the Sarkis ionization efficiency (quenching) model~\cite{sarkis2022} and Ramanathan charge production model~\cite{Ramanathan:2020fwm}. 
    In both cases, we assume perfect backing detector with full azimuthal coverage. 
    {\bf Left:} for higher neutron energies and wide angles, the contribution from the inner shell~\cite{Ibe:2017yqa} is distinct above the elastic peak. {\bf Right:} for low neutron energies and shallow angles, the contribution from the valence band~\cite{Knapen:2021bwg} separates from the elastic peak. 
    }
    \label{fig:Eion_spec}
\end{figure*} 

To predict the number of electron-hole pairs $n_e$ as a function of $E_{ion}$, we use the charge production model presented in Ref.~\cite{Ramanathan:2020fwm}. 
This is a data-driven model of impact ionization in silicon that more accurately models the response for low $n_e$ than a model of Fano statistics alone (i.e. Ref.~\cite{fano}).  Ref.~\cite{Ramanathan:2020fwm} provides a set of functions $p_{n_e}(E_{ion})$ for the probability of producing $n_e$ pairs for energy deposit $E_{ion}$. 
Thus, to compute measured ionization rates as a function of angle, we integrate Eq.~\eqref{eq:result} against $p_{n_e}$ to find the differential angular probability of Migdal events binned in $n_e$,
\begin{equation}
    \frac{d P_{n_e}}{d \cos \theta_n} = \int dE_{ion}  \hspace{1mm} p_{n_e}(E_{ion})  \frac{d^2 P_{M}}{d \cos \theta_n d E_{ion}}. 
\end{equation}
The inherent widths of the $p_{n_e}$ leads to a smearing effect that can affect our signal (even before considering experimental factors, such as non-monochromaticity of the beam).
We show two examples of observable spectra in Fig.~\ref{fig:Eion_spec} for different choices of $E_n$ and $\theta_n$, where we decompose the spectral contributions to the rate from elastic, valence band ME, and inner shell ME scatters.


Fig.~\ref{fig:Eion_spec} illustrates two possible strategies for calibrating the ME in the correct kinematic regime. 
The $q^2$ scaling of the Migdal probabilities translates to an enhancement approximately proportional to $E_n (1 - \cos \theta_n)$.
Larger momentum transfers (which lead to larger nuclear recoil energies), achieved either by raising the neutron beam energy or by looking at a larger scattering angle, will therefore give a higher rate of Migdal events. 
However, to avoid the aforementioned difficulties of the inherent smearing due to the Fano statistics, it is important to keep the nuclear recoil energy scale small enough that the elastic spectrum does not smear too much into the Migdal tail.
Thus, the first experimental strategy is to target setups that balance the $q^2$ rate enhancement with low recoil energy, in order to clearly isolate the high-side Migdal rate tail. 
As can be seen in the left plot of  Fig.~\ref{fig:Eion_spec}, this strategy is particularly useful for calibrating Eq.~\eqref{eq:ibe}, the ME contribution for inner shell electrons, but care must be taken not to increase the neutron energy and scattering angle outside of the kinematic regime of interest (see Appendix~\ref{app:softlimit}). 

The second strategy is to employ low-energy neutrons scattering at low angles such that the quenched nuclear recoils are too small to produce \textit{any} secondary ionization, effectively eliminating the observable elastic contribution (the second term of Eq.~\eqref{eq:quenching}). 
This strategy is challenging in that it involves novel neutron source development, but is able to calibrate Eq.~\eqref{eq:Lin}, the ME contribution from valence electrons independent of other contributions, as shown in the right plot of Fig.~\ref{fig:Eion_spec}.  
Since sub-GeV DM will typically only produce single- to few-electron events, this setup more closely mimics what a sub-GeV DM signal would look like in a single-electron threshold charge detector. 

In a real experiment, no neutron beam will be perfectly monochromatic, such that contamination from higher-energy neutrons and gamma backgrounds must be taken carefully into account through validated simulations. 
There are a number of common neutron sources and methods that are implemented in the lab, each of which can be turned into a fairly monochromatic beam with careful application. These include deuterium-deuterium (D-D) and deuterium-tritium (D-T) generators~\cite{ddgenerator}, proton accelerators incident on a $^7$Li~\cite{Joshi:2014oda} or $^{51}$V~\cite{PhysRev.100.167} target, and photoneutron sources that exploit the $^9$Be disintegration threshold of 1.67~MeV~\cite{Collar:2013xva,Robinson:2016kxn}.
Each of these options comes with its own advantages and disadvantages, so we will emphasize that the fairly rare probability of a Migdal scatter, even in an ideal setup, necessitates (a) a low-background environment, as is typically achieved with significant overburden, thus complicating the use of proton accelerators, and (b) a high flux of low-energy neutrons, which can be achieved with either photoneutron sources or moderated D-D (or D-T) generators. 
Collimation in both of these cases is achieved through robust $4\pi$ shielding minus a small beam hole, typically around 1~cm in diameter (dependent on detector size).

Because of logistical challenges associated with having a sufficiently high-activity gamma source to produce a high flux of neutrons from a photoneutron source, we will focus the rest of the Letter on using a D-D neutron generator, as is employed by NEXUS. 
A D-D generator leverages fusion reactions to generate isotropic 2.5~MeV neutrons without any primary gamma backgrounds~\cite{ddgenerator} (although secondary gammas will be produced by neutrons interacting in surrounding shielding materials). 
Using clever application of ``filters," it is possible to prune, or even adjust, the beam energy spectrum by exploiting anti-resonances in the neutron scattering cross section~\cite{BARBEAU2007385}.
Filters have the added advantage of removing unwanted secondary gamma backgrounds from neutron interactions in the shield materials and any primary x-rays produced by the generator, which can be shielded by even a small amount of material. 
Of note, prominent anti-resonances in iron and scandium can be used to select 24~keV~\cite{FeFilter} and 2~keV~\cite{ScFilter} neutrons, respectively. 
A downside of using filters to select an optimal beam energy is a substantial reduction in neutron flux, thus requiring longer exposures, hotter sources, and lower ambient backgrounds. 
Another option for reducing the neutron energy is to employ neutron reflectors~\cite{Verbus:2016sgw}, but this would require more substantial modification to the NEXUS setup, and so is not the focus of this study. 
Lower-energy neutron beams also mandate progress in low-energy neutron backing detectors, which is an active area of study~\cite{Biekert:2022qtk} but outside the scope of this Letter. 

As a schematic setup, the NEXUS facility at Fermilab is designed to provide a D-D generator neutron beam incident on a 10~mK, single-electron resolution detector (e.g. SuperCDMS HVeV~\cite{Romani:2017iwi,SuperCDMS:2020ymb,Ren:2020gaq,SuperCDMS:2022zmd}) in a $\sim$100~cts/kg/day/keV radiation environment. 
Crucially, the chosen detector should be thin compared to the mean free path in silicon of $\sim$10~cm for a $<$50~keV neutron (for which the cross section is constant~\cite{endf}), to ensure that neutrons scatter only once on average and thus that angular smearing from multiple scattering is not a concern. The NEXUS D-D generator (Adelphi model DD108) produces up to $\sim$10$^9$ neutrons/s isotropically, with a collimated $\sim$10$^3$ neutrons/s rate incident on a $\sim$1~cm$^2$ detector area. 
Filters should be able to modulate a higher-energy neutron source (such as the D-D generator) down to the respective anti-resonance with roughly 10$\%$ energy width, minimal higher-energy contamination, and a $\sim$10$^3$ reduction in overall flux. 
This means that, with minimal modification, NEXUS could produce a filtered keV-scale neutron beam with $\sim$1~neutron/s incident on a single-electron threshold semiconductor detector with a custom backing array in a low-background environment; to the best of our knowledge, no other such facility currently exists. 

In such a setup with a series of $\Delta\theta_n \approx 10^{\circ}$ wide backing arrays at different angles (including at $\theta_n \pm 5^{\circ}$ around the central angles shown in Fig.~\ref{fig:Eion_spec}), one would expect to see only a handful of neutron-induced Migdal events with roughly one month of exposure in the case of the 24~keV beam. 
This should be sufficient to calibrate the normalization of the electronic matrix element in Eq.~\eqref{eq:ibe}, but upgrades to increase the rate would be required to fully reconstruct it. 
Meanwhile, the 2~keV beam setup requires more exposure than is practical without a more substantial upgrade to NEXUS in order to calibrate even the normalization of the matrix element in Eq.~\eqref{eq:Lin}.  
To accomplish this, the rate could be increased with a hotter D-D (or D-T) neutron source or by deploying multiple silicon detectors in the beam, but each of these improvements comes with trade-offs and complications. 
One of the biggest challenges in any of these setups would be to sufficiently eliminate higher-energy neutron contamination in the beam from the energy region of interest, which can hopefully be achieved through careful angular tagging. 


In this work we have extended previous calculations of the ME to study the angular distributions of the neutron-induced ME in silicon.
These results can also be applied to an atomic target calibration by using Eq.~\eqref{eq:ibe} for the valence shell as well as the inner shells. 
We have demonstrated that Migdal scatters leave a distinct pattern in ionization measurements at fixed angles, providing a clear experimental target for calibration studies. 
We further emphasize that inherent spreading in the energy resolution in silicon strongly motivates the use of lower-energy neutrons and angular selection for a clean measurement. 
Lower-energy neutrons are also better kinematically tuned to mimic sub-GeV DM scattering, allowing direct calibration of ME probabilities in the kinematic regime of interest. 
Practical applications of this work will need to account for detector-specific backgrounds and non-ideal beam effects in their design, as well as the backgrounds from inelastic nuclear scattering. 
This work lays out necessary steps toward the calibration of the ME with neutrons in silicon (and germanium), which will be crucial to validate both existing (e.g. EDELWEISS~\cite{EDELWEISS:2019vjv} CDEX~\cite{CDEX:2019hzn}, SuperCDMS~\cite{SuperCDMS:2022kgp}, DAMIC at SNOLAB~\cite{DAMIC:2019dcn}, and SENSEI~\cite{SENSEI:2020dpa}) and next-generation (e.g. SENSEI at SNOLAB,  DAMIC-M~\cite{Castello-Mor:2020jhd,Settimo:2020cbq}, Oscura~\cite{2022arXiv220210518A}, and SuperCDMS SNOLAB~\cite{SuperCDMS:2016wui}) limits on sub-GeV DM-nuclear scattering. 
\\

\begin{acknowledgments}
We thank Alvaro Chavarria, Juan Collar, Juan Estrada, Gordan Krnjaic, Noah Kurinsky, Brian Lenardo, Junsong Lin, Pat Lukens, Daniel McKinsey, Karthik Ramanathan, Javier Tiffenberg, Belina von Krosigk, and Kevin Zhang for useful conversations related to the content of this Letter.
We are especially grateful to Enectal\'i Figueroa-Feliciano, Scott Hertel, Lauren Hsu, Tongyan Lin, Youssef Sarkis, Jingke Xu, and Tien-Tien Yu for their feedback on early drafts of this analysis.
The work of H.D.~and Y.K.~is supported in part by DOE Award No. DE-SC0015655. 
R.E.~and D.A.~are supported by DoE Award No. DE-SC0009854 and Simons Investigator in Physics Grant No. 623940.  R.E.~also acknowledges support from the US-Israel Binational Science Foundation Grant No.~2016153 and the Heising-Simons Foundation Grant No.~79921.  
D.B.~is supported by the U.S.~Department of Energy, Office of Science, National Quantum Information Science Research Centers, Quantum Science Center. 
Fermilab is operated by Fermi Research Alliance, LLC, under Contract No.~DE-AC02-07CH11359 with the U.S. Department of Energy. 
\end{acknowledgments}

%

\include{sm_main}

\end{document}

%% file: sm_main.tex
\begin{widetext}

\begin{appendix}

\section{Inelastic Scattering Kinematics in the Soft Limit}
\label{app:chargevar}

In this Appendix, we derive the kinematics for inelastic 2-body scattering in the soft and free-ion limits, where the initial-state nucleus is at rest and the electron system takes energy $\omega$ but no momentum. In particular, the derivation in the lab frame is necessary to keep track of the scattered neutron angle; previous derivations from e.g.~Ref.~\cite{Ibe:2017yqa} are in the center-of-mass frame, with all angular dependence integrated out, whereas we want to preserve the angular dependence in the lab frame.

In the lab frame, and under the assumptions of the soft and free-ion limits, energy conservation gives
\begin{equation}
    \frac{1}{2m_n} (|\vec{p}_i|^2 - |\vec{p}_f|^2 ) = E_r + \omega \,,
\end{equation}
where $\vec{p}_i$ and $\vec{p}_f$ are the initial and final momentum of the neutron, respectively, while momentum conservation gives
\begin{equation}
\begin{split}
    |\vec{q}|^{2} &= |\vec{p}_i - \vec{p}_f |^2  = |\vec{p}_i|^2 + |\vec{p}_f|^2 - 2|\vec{p}_i||\vec{p}_f|\cos\theta_n \\
    &= |\vec{q_N}|^2 = 2 m_N E_r \,,
\end{split}
\end{equation}
since the momentum $\vec{q}$ transferred to the target goes entirely to the recoiling nucleus, which gets momentum $\vec{q}_N = \vec{q}$. We can thus rewrite the energy conservation equation as
\begin{equation}
    2m_n (E_n - E_r - \omega)  = |\vec{p}_f|^2 = 2 m_N E_r - 2 m_n E_n + 2|\vec{p}_i||\vec{p}_f|\cos\theta_n \,.
\end{equation}
Rearranging terms and plugging in for $|\vec{p}_f|$, we find
\begin{equation}
\begin{split}
    E_n - E_r - \omega  &=  \frac{m_N}{m_n} E_r - E_n + \frac{\sqrt{2 m_n E_n} \sqrt{2 m_n (E_n - E_r - \omega)} \cos\theta_n}{m_n} \\
   \implies  E_n \left( 1 - \frac{E_r + \omega}{E_n} \right) &= \frac{m_N}{m_n} E_r - E_n + 2 E_n \sqrt{ 1 - \frac{E_r + \omega}{E_n}} \cos\theta_n \,,
\end{split}
\end{equation}
Finally, we can simplify to
\begin{equation}
    \cos\theta_n = \frac{ E_n \left( 2 - \frac{E_r + \omega}{E_n} \right) - \frac{m_N}{m_n} E_r}{2 E_n \sqrt{ 1 - \frac{E_r + \omega}{E_n}}} 
    \, .
\end{equation}
Inverting this to solve for $E_r$ yields a quadratic equation with solution
\begin{equation}\label{eq:ervstheta}
\begin{split}
E_r &= \frac{2 E_n m_n }{(m_n+m_N)^2} \left(m_n \sin^2 \theta_n + m_N - \cos \theta_n \sqrt{m_N^2 - m_n^2 \sin^2 \theta_n - \frac{m_N(m_n + m_N) \omega}{E_n}}\right) - \frac{m_n \omega}{m_n + m_N} \\
&= \frac{2 E_n \mu^2}{m_n m_N} \left( \frac{m_n}{m_N} \sin^2\theta_n + 1 - \cos\theta_n \sqrt{ \frac{m_n^2}{m_N^2} (\cos^2\theta_n -1) + 1 - \frac{m_n \omega}{\mu E_n} } \right) - \frac{\mu \omega}{m_N}.
\end{split}
\end{equation}
In principle, there are two roots, but for $m_n < m_N$ (as is always the case in our setup) the second root is spurious. The choice of root is consistent with the limit $\omega \to 0$, where our expression reduces to standard textbook results for $2 \to 2$ elastic scattering of unequal masses~(e.g.~\cite{thornton2021classical}):
\begin{equation}
E_r(\omega = 0) =  E_n \left(1 - \frac{m_n^2}{(m_n + m_N)^2 }\left( \cos \theta_n + \sqrt{\frac{m_N^2}{m_n^2} - \sin^2 \theta_n}\right)^2 \right).
\end{equation}
We note that many of these formulas simplify somewhat in the center-of-mass frame, where the neutron scattering angle $\theta_n' = \theta_n + \mathcal{O}(m_n/m_N)$ is almost identical to the lab frame for $m_n \ll m_N$. However, for small angles the corrections are significant, so we work in the lab frame for consistency. We note that in this limit, these results are analogous to the more simplified center-of-mass Eq.~(94) from Ref.~\cite{Ibe:2017yqa}. 

Finally, in order to translate the standard result into angular coordinates, we must take the derivative of Eq.~\eqref{eq:ervstheta} with respect to $\cos\theta_n$, which yields
\begin{equation}
\begin{split}
 \left | \frac{dE_r}{d\cos\theta_n} \right |=&~  \frac{2 E_n m_n \left (m_n \cos \theta_n + \sqrt{m_N^2 - m_n^2 \sin^2 \theta_n - \frac{m_N(m_n + m_N) \omega}{E_n}}\right )^2}{(m_n + m_N)^2 \sqrt{m_N^2 - m_n^2 \sin^2 \theta_n - \frac{m_N(m_n + m_N) \omega}{E_n}}} \\ =&~ 
 \frac{2 E_n \mu^2 \left (\frac{m_n}{m_N} \cos \theta_n + \sqrt{ \frac{m_n^2}{m_N^2} (\cos^2 \theta_n -1) + 1 - \frac{m_n \omega}{\mu E_n}}\right )^2}{ m_n m_N \sqrt{ \frac{m_n^2}{m_N^2} (\cos^2 \theta_n - 1) + 1 - \frac{m_n \omega}{\mu E_n}}}\,.
\end{split}
\label{eq:dEdcostheta}
\end{equation}

\section{Angular Dependence of the Migdal Effect for Semiconductors in the Soft Limit}
\label{app:softlimit}

In this Appendix, we adapt the formalism of Ref.~\cite{Knapen:2020aky} to derive the angular spectrum of the scattered neutron in the soft limit, restoring the dependence on the momentum $\vec{k}$ transferred to the electronic system, and keeping careful track of any assumptions or approximations made along the way. We begin with the general expression for the electronic energy spectrum in the soft limit, Eq.~(A33) in Ref.~\cite{Knapen:2020aky}:
\beq
\begin{aligned}
\frac{d\sigma}{d\omega} = C_\chi \int\frac{d^3 \vec{q}}{(2\pi)^3}&\int\frac{d^3\vpf}{(2\pi)^3}|F(\vpi-\vpf-\vec{q})|^2 \delta\left( E_i-E_f-\omega-\tfrac{q^2}{2m_N}\right)\\
\times&\int\frac{d^3\vec{k}}{(2\pi)^3}\sum_{\vec{K}}Z_{\text{ion}}^2(|\vec{k}+\vec{K}|)\frac{\text{Im}|-\epsilon_{KK}^{-1}|}{|\vec{k}+\vec{K}|^2}\frac{|\vec{q}\cdot(\vec{k}+\vec{K})|^2}{\omega^4 m_N^2}\,,
\label{eq:startsoft}
\end{aligned}
\eeq
where $\vec{K}$ is a reciprocal lattice vector, $F$ is a form factor parametrizing the zero-point momentum spread of the initial-state nucleus, and $\epsilon$ is the dielectric function of the target. As in Appendix~\ref{app:chargevar} above, the momentum transferred to the target $\vec{q}$ is equal to the momentum of the recoiling nucleus $\vec{q}_N$ in the soft limit, so we use $\vec{q}$ instead of $\vec{q}_N$ to maintain the distinction with the full calculation outside the soft limit in Appendix~\ref{app:nosoft} below. In the case of DM scattering, we typically integrate over the unobserved DM momentum $\vpf$, but for neutron scattering we want to keep $\vpf$ and integrate over the unobserved nuclear recoil momentum $\vec{q}$. The prefactor also changes, with
\beq \label{eq:dmton}
C_\chi \equiv \frac{8\pi\alpha}{v_\chi} \left( \frac{2\pi b_\chi}{\mu_{\chi N}}\right)^2 \rightarrow C_n \equiv \frac{8\pi\alpha}{v_n}\left(\frac{2\pi b_n}{m_n}\right)^2\,, 
\eeq
where $v_n$ is the initial neutron velocity and $b_n$ is the neutron scattering length in silicon. Note that the convention in the neutron scattering literature is typically to define $b_n$ such that the neutron mass rather than the reduced mass appears in Eq.~(\ref{eq:dmton}); for $^{28}{\rm Si}$ the corresponding value of $b_n$ is 4.1 fm~\cite{endf}. For simplicity of notation, we will often abbreviate $\vec{k} + \vec{K} \equiv \vec{k}'$ and $\int d^3 \vec{k} \sum_{\vec{K}} \to \int d^3 \vec{k}'$, since the lattice structure will not be essential to our arguments. We will also write $\mathcal{W}(\vec{k'}, \omega) \equiv \text{Im}(-\epsilon^{-1}(\vec{k'}, \omega))$ for the ELF.

Following Ref.~\cite{Knapen:2020aky}, we make the free-ion approximation where $F$ can be replaced by a momentum-conserving delta function,
\beq
|F(\vpi-\vpf-\vec{q})|^2 \to (2\pi)^3 \delta(\vpi - \vpf - \vec{q})\,.
\eeq
The form factor $F^2$ is a Gaussian with width $q_0 = \sqrt{2 m_N \omega_{ph}} \simeq 56 \ {\rm keV}$ in Si, where $\omega_{ph} \simeq 60 \ {\rm meV}$ is a typical optical phonon energy. Note that the impulse approximation already requires $q \gg q_0$, so the spread in $F^2$ will typically not induce a large deviation from exact momentum conservation when the impulse approximation is satisfied. The impulse approximation will be valid for the kinematic regime we consider ($p_i \gg q_0$, $\theta_n$ not too small), but may fail for small neutron energies and/or very forward scattering. However, see Refs.~\cite{Liang:2022xbu,Berghaus:2022pbu}, which demonstrate that the impulse approximation may be extended below its nominal regime of validity and coincides almost exactly with a full treatment using the phonon density of states. 

Performing the $\vec{q}$ integral in Eq.~(\ref{eq:startsoft}) using the delta function amounts to the replacement $\vec{q}=\vpi-\vpf$. This leaves
\beq
\frac{d\sigma}{d\omega} = C_n \int\frac{d^3\vpf}{(2\pi)^3} \delta\left( E_i-E_f-\omega-\tfrac{(\vpi-\vpf)^2}{2m_N}\right)\int\frac{d^3\vec{k}'}{(2\pi)^3}Z_{\text{ion}}^2(k')\frac{\mathcal{W}(\vec{k'}, \omega)}{|\vec{k}'|^2}\frac{|(\vpi-\vpf)\cdot\vec{k}'|^2}{\omega^4 m_N^2}\,.
\eeq
The radial part of the $\vpf$ integral can be performed using the energy-conserving delta function, for which the algebra is equivalent to the derivation in Appendix~\ref{app:chargevar}. The azimuthal integral is trivial and gives a factor of $2\pi$, leaving
\beq
\label{eq:withdots}
\frac{d\sigma}{d\cos\theta_n d\omega } = C_n \int\frac{d^3\vec{k}'}{(2\pi)^5}Z_{\text{ion}}^2(k')\frac{\mathcal{W}(\vec{k'}, \omega)}{k'^2}\left(\frac{(p_f^+)^2|(\vpi-\vpf^+)\cdot\vec{k}'|^2+(p_f^-)^2|(\vpi-\vpf^-)\cdot\vec{k}'|^2}{\omega^4 m_N^2\sqrt{\frac{p_i^2}{m_N^2}(\cos^2\theta_n-1)+\frac{p_i^2}{m_n^2}-2\omega\frac{m_n+m_N}{m_n m_N}}}\right)\,,
\eeq
where 
\beq
p_f^\pm = \frac{m_np_i\cos\theta_n}{m_n+m_N}\pm\sqrt{\left(\frac{m_np_i\cos\theta_n}{m_n+m_N}\right)^2-\frac{2m_nm_N\omega+p_i^2(m_n-m_N)}{m_n+m_N}}\,.
\label{eq:pfpm}
\eeq
The square root in the definition of $p_f$ can, in principle, restrict the range of scattering angles:
\beq
\cos^2\theta_n \geq\frac{m_N^2}{m_n^2}\left(\frac{2m_n\omega}{p_i^2}-1\right)+\frac{2m_N\omega}{p_i^2}+1\,.
\eeq
The kinematic threshold where scattering is forbidden occurs when the right-hand side is greater than 1. For $\omega \ll E_n$ and $m_n < m_N$, which will always be the case for the kinematics we consider, the right-hand side is negative and there is no angular restriction, but the $p_f^-$ solution is negative and therefore spurious. We will thus relabel $p_f^+ \to p_f$. There is a very narrow range of energies close to threshold, $\omega \in \left [ \frac{E_n(m_N^2 - m_n^2)}{m_n^2 + m_N^2}, \frac{E_n m_N^2}{m_n^2 + m_N^2} \right ] \simeq E_n$, where both roots are allowed. This is an extremely fine-tuned kinematical region, with the difference between the lower and upper boundaries being $\Delta \omega/E_n = \frac{m_n^2}{m_n^2 + m_N^2} \simeq 10^{-3}$ for Si, and thus it is outside the regime of relevance for these studies (both because our proposed neutron source does not have this precision on the initial energy, and because we are never considering order-1 fractions of the initial energy taken by the electrons). However, it may be relevant for neutron scattering on very light targets such as helium.

As a check on these results, consider the elastic limit $\omega\rightarrow0$. The angular restriction from the square root is
\beq
\cos^2\theta_n \geq 1-\frac{m_N^2}{m_n^2} \qquad (\omega \to 0)\,,
\eeq
which is always satisfied for any $\theta$ as long as $m_N \geq m_n$. The solution for $p_f$ becomes
\beq
p_f=\frac{p_i}{m_n+m_N}\left( m_n\cos\theta + \sqrt{m_N^2-m_n^2\sin^2\theta}\right) \qquad (\omega \to 0)\,,
\eeq
which recovers the classical elastic scattering results.

To simplify the dot products in Eq.~(\ref{eq:withdots}), note from the original form of the energy delta function that 
\beq
\frac{(\vec{p}_i-\vec{p}_f)^2}{m_N}=\frac{p_i^2}{m_n}-\frac{p_f^2}{m_n}-2\omega\,,
\label{eq:qsqenergy}
\eeq
so
\beq
|(\vpi-\vpf)\cdot\vec{k}'|^2=(|\vpi-\vpf||\vec{k}'|\cos\theta_k)^2=\frac{m_N}{m_n}\left( p_i^2-p_f^2-2m_n\omega\right) k'^2\cos^2\theta_k\,,
\eeq
where $\theta_k$ is the angle between the momentum transfer $\vpi-\vpf$ and the momentum in the electron system $\vec{k}'$. Combining everything, we now have
\beq
\frac{d\sigma}{d\cos\theta_n d\omega } = C_n \int  \frac{d^3\vec{k}'}{(2\pi)^5} Z_{\text{ion}}^2(k')\mathcal{W}(\vec{k'}, \omega) 
\frac{p_f^2(p_i^2- p_f^2 - 2m_n\omega)\cos^2 \theta_k}{\omega^4 m_nm_N\sqrt{\frac{p_i^2}{m_N^2}(\cos^2\theta_n-1)+\frac{p_i^2}{m_n^2}-2\omega\frac{m_n+m_N}{m_nm_N}}}\,.
\eeq
At this point, we are able to perform the remaining angular integrals if we assume isotropy of the target, such that $Z_{\text{ion}}^2$ and $\mathcal{W}$ depend only on $|\vec{k'}|$. This assumption does not hold exactly for any lattice structure, but is likely a reasonable approximation for the highly-symmetric diamond cubic crystal structure of silicon (or germanium). Assuming isotropy, the azimuthal integral trivially gives a factor of $2\pi$, and treating $\cos\theta_k$ as the polar angle of the $\vec{k}'$ integral, we pick up a factor of $\int d\cos \theta_k \cos^2 \theta_k= \frac{2}{3}$.
Thus, we are left with
\beq
\label{eq:almostdone}
\frac{d\sigma}{d\cos\theta_n d\omega } = C_n \frac{2p_f^2}{3\omega^4 m_nm_N}\frac{p_i^2-p_f^2-2m_n\omega}{\sqrt{\frac{p_i^2}{m_N^2}(\cos^2\theta_n-1)+\frac{p_i^2}{m_n^2}-2\omega\frac{m_n+m_N}{m_nm_N}}}\int\frac{dk'k'^2}{(2\pi)^4}Z_{\text{ion}}^2(k')\mathcal{W}(k', \omega)\,.
\eeq

Eq.~(\ref{eq:almostdone}) shows that, under the assumptions of isotropy and the soft limit, the only kinematic dependence of the integrand is carried by $\omega$, and thus the Migdal rate factorizes as claimed in the main text. Indeed, the integral in Eq.~(\ref{eq:almostdone}) is proportional to the electronic spectrum, Eq.(\ref{eq:Lin}) (repeated here for convenience),
\begin{equation}
\frac{d\widetilde{P}_e}{d\omega} = \frac{4 \alpha }{\omega^4 m_N^2} \int \frac{d^3 k'}{(2\pi)^3} Z_{\rm ion}^2(k')(\hat{q} \cdot \hat{k}')^2 \, \mathcal{W}(k', \omega) \approx \frac{8 \alpha }{3 \omega^4 m_N^2} \int \frac{dk' k'^2}{(2\pi)^2} Z_{\rm ion}^2(k') \,, \mathcal{W}(k', \omega)
\end{equation}
where in the second equality we have used the assumed isotropy of the ELF to integrate over the angles. Restoring the prefactor $C_n$, this gives the desired factorization,
\begin{equation}
\frac{d\sigma}{d\cos\theta_n d\omega } = \frac{2 \pi b_n^2 m_N}{m_n^3 v_n} \frac{p_f^2 (p_i^2-p_f^2-2m_n\omega)}{\sqrt{\frac{p_i^2}{m_N^2}(\cos^2\theta_n-1)+\frac{p_i^2}{m_n^2}-2\omega\frac{m_n+m_N}{m_nm_N}}} \frac{d\widetilde{P}_e}{d\omega}\,,
\end{equation}
where $d\widetilde{P}_e/d\omega$ depends only on $\omega$ and may be calculated  using the \texttt{DarkELF} code package~\cite{Knapen:2021bwg} independent of the neutron scattering experimental parameters.

To convert the cross section to a probability per neutron $P_M$, we use the relation
\begin{equation}
    \frac{dP_M}{d \cos \theta_n d \omega} = \frac{N_0 \rho_T L}{A_N} \frac{d\sigma}{d\cos\theta_n d\omega }\,,
\end{equation}
where $\rho_T$ is the mass density of the target, $L$ is the thickness of the target, $N_0$ is Avogadro's number, and $A_N \approx m_N/m_n$ is the atomic mass number of the target. This expression is valid when $L$ is much less than the neutron mean free path, which (as discussed in the main text) is necessary to prevent angular smearing from multiple scattering. Using the definition of the elastic cross section in terms of the scattering length, $\sigma_{el} \equiv 4\pi b_n^2$, and substituting $p_i^2 = 2 m_n E_n$ and Eq.~(\ref{eq:pfpm}) for $p_f$, gives Eq.~(\ref{eq:lab}) in the main text which is an explicit expression for the angular spectrum in terms of the experimental variables $E_n$, $\omega$, and $\cos \theta_n$. 

As a final check on our results, we recover the original form of the Migdal rate as an energy spectrum, Eq.~(\ref{eq:RMGeneral}) as follows. First, use Eq.~(\ref{eq:qsqenergy}) to replace $p_i^2 - p_f^2 - 2 m_n \omega$ in the numerator with $q^2 (m_n/m_N)$. Converting the scattering probability to a rate $R_M$ by multiplying by $\Phi A$, where $\Phi$ is neutron flux in neutrons/cm$^2$/s and $A$ is target area, gives
\begin{equation}
\frac{dR_M}{d \cos \theta_n d \omega} = \frac{N_0 \rho_T L \Phi A}{A_N}\frac{\sigma_{el}}{2}\frac{\mu^2}{m_n^2}\frac{\left(\frac{m_n}{m_N}\cos \theta_n + \beta\right)^2}{\beta} q^2 \frac{d\widetilde{P}_e}{d\omega}\,,
\end{equation}
where $\beta$ is defined in Eq.~(\ref{eq:betadef}). We can rewrite this using the Jacobian computed in Eq.~(\ref{eq:dEdcostheta}),
\begin{align}
    \frac{dR_M}{d \cos \theta_n d \omega} & = \frac{N_0 \rho_T V \Phi}{A_N}\frac{\sigma_{el} m_N}{4 m_n E_n}\left | \frac{dE_r}{d \cos \theta_n} \right | q^2 \frac{d\widetilde{P}_e}{d\omega} \nonumber \\
    & = \frac{dR_{el}}{dE_r} \left | \frac{dE_r}{d\cos\theta_n} \right | q^2 \frac{d\widetilde{P}_e}{d\omega}\,,
\label{eq:SpectrumSoft}
    \end{align}
where in the first equality we have replaced $A L$ with $V$, the volume of the target. The second equality follows from the standard formulas for elastic nuclear recoil because we have defined $\sigma_{el}$ using the neutron scattering convention where the target is treated as infinitely heavy, $\mu \to m_n$. 

Fig.~\ref{fig:SpectrumAppendix} shows the spectrum resulting from Eq.~(\ref{eq:result}) for a broader range of kinematics than shown in the main text. For nuclear quenching, we consider the Sarkis model~\cite{Sarkis:2020soy} and the Lindhard model~\cite{osti_4701226} as lower and upper bounds on the quenching factor, respectively. At the highest neutron energies and wide scattering angles, the soft-limit approximation clearly fails because $q$ is too large, such that the scaling with $q^2$ is unphysical and the Migdal rate appears enhanced compared to the elastic scattering rate. We will discuss other failures of the soft limit in Appendix~\ref{app:nosoft} below.

\begin{figure}[!t]
\includegraphics[width=0.49\textwidth]{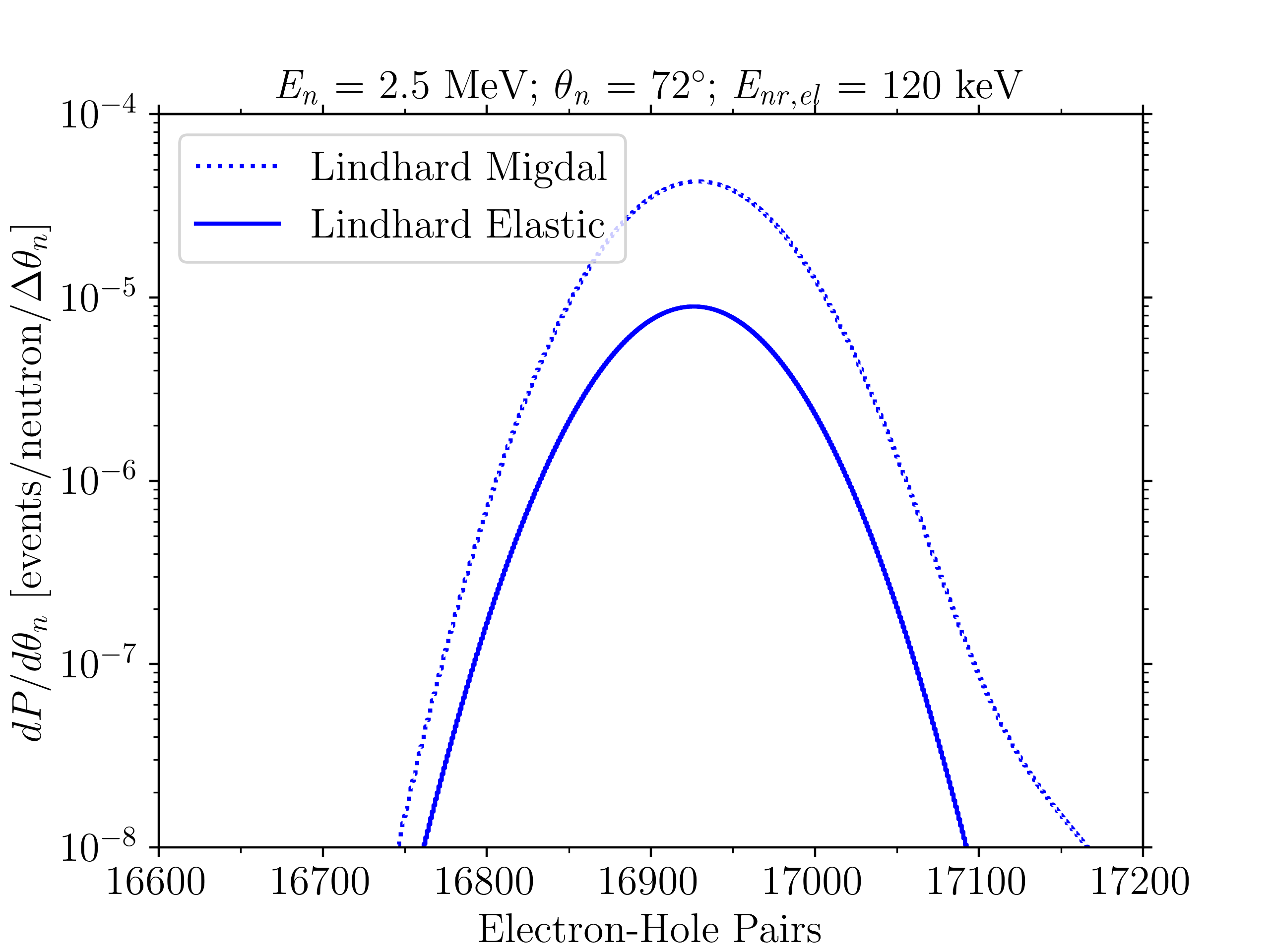}~
\includegraphics[width=0.49\textwidth]{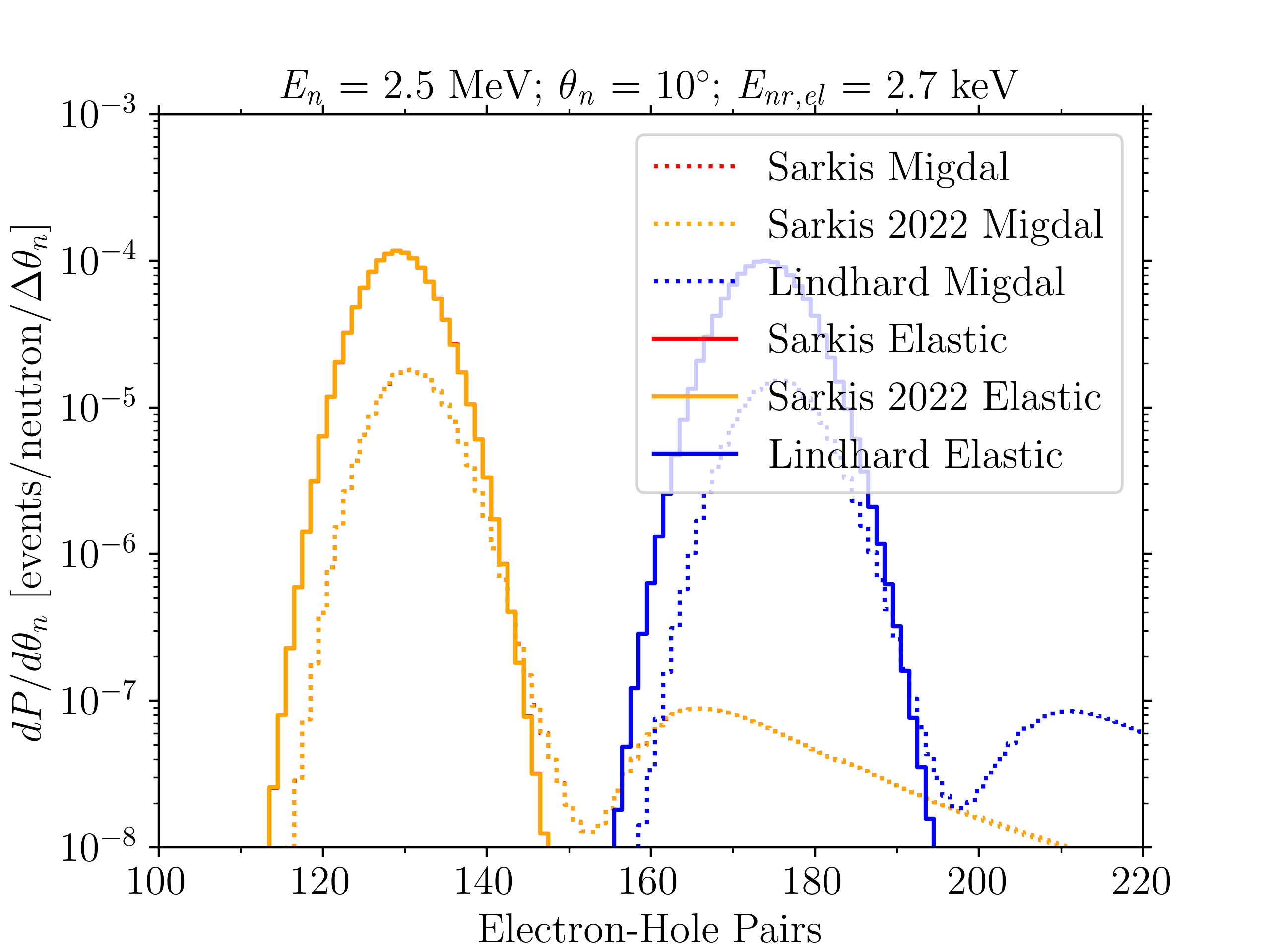}
\includegraphics[width=0.49\textwidth]{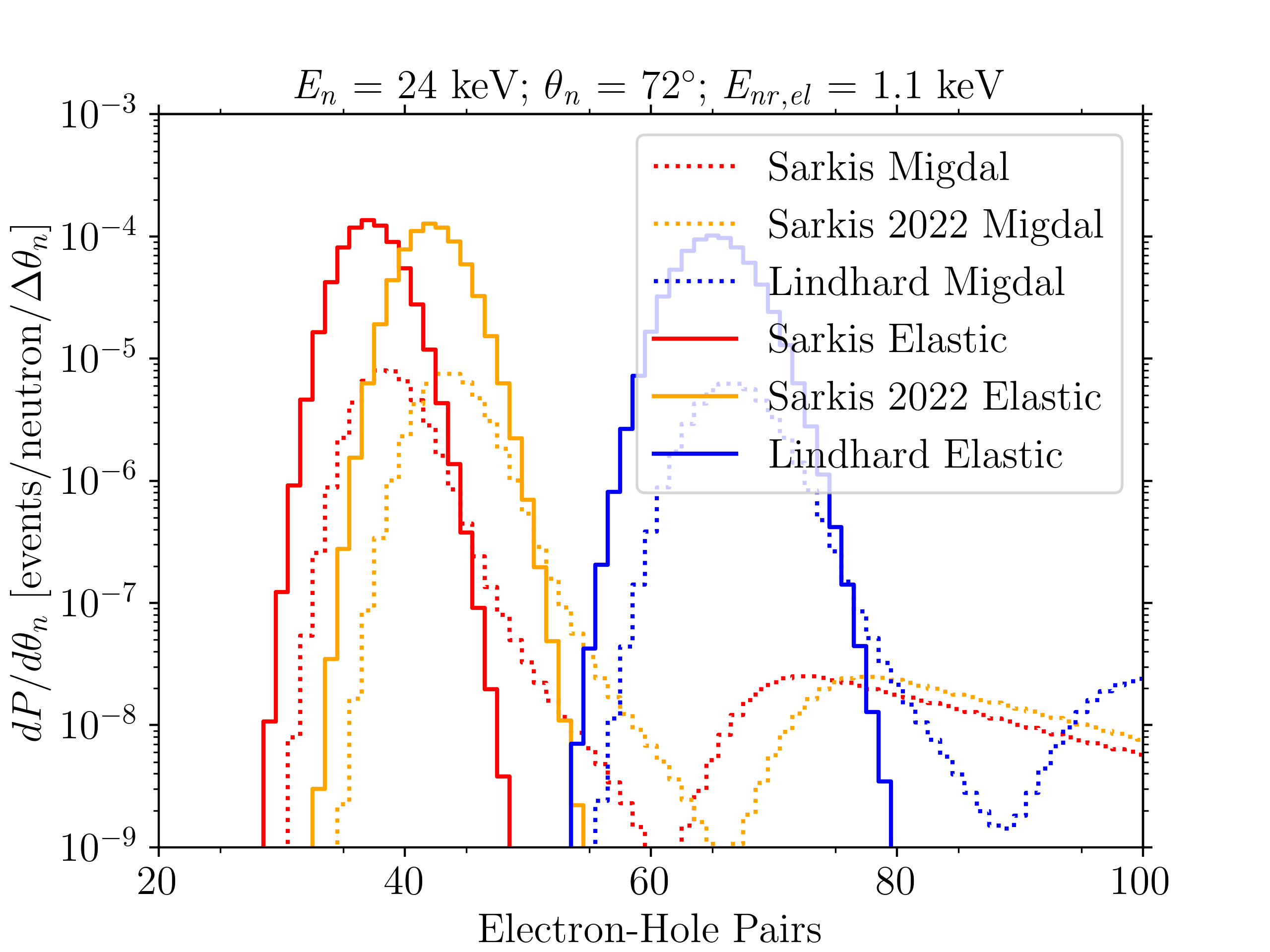}~
\includegraphics[width=0.49\textwidth]{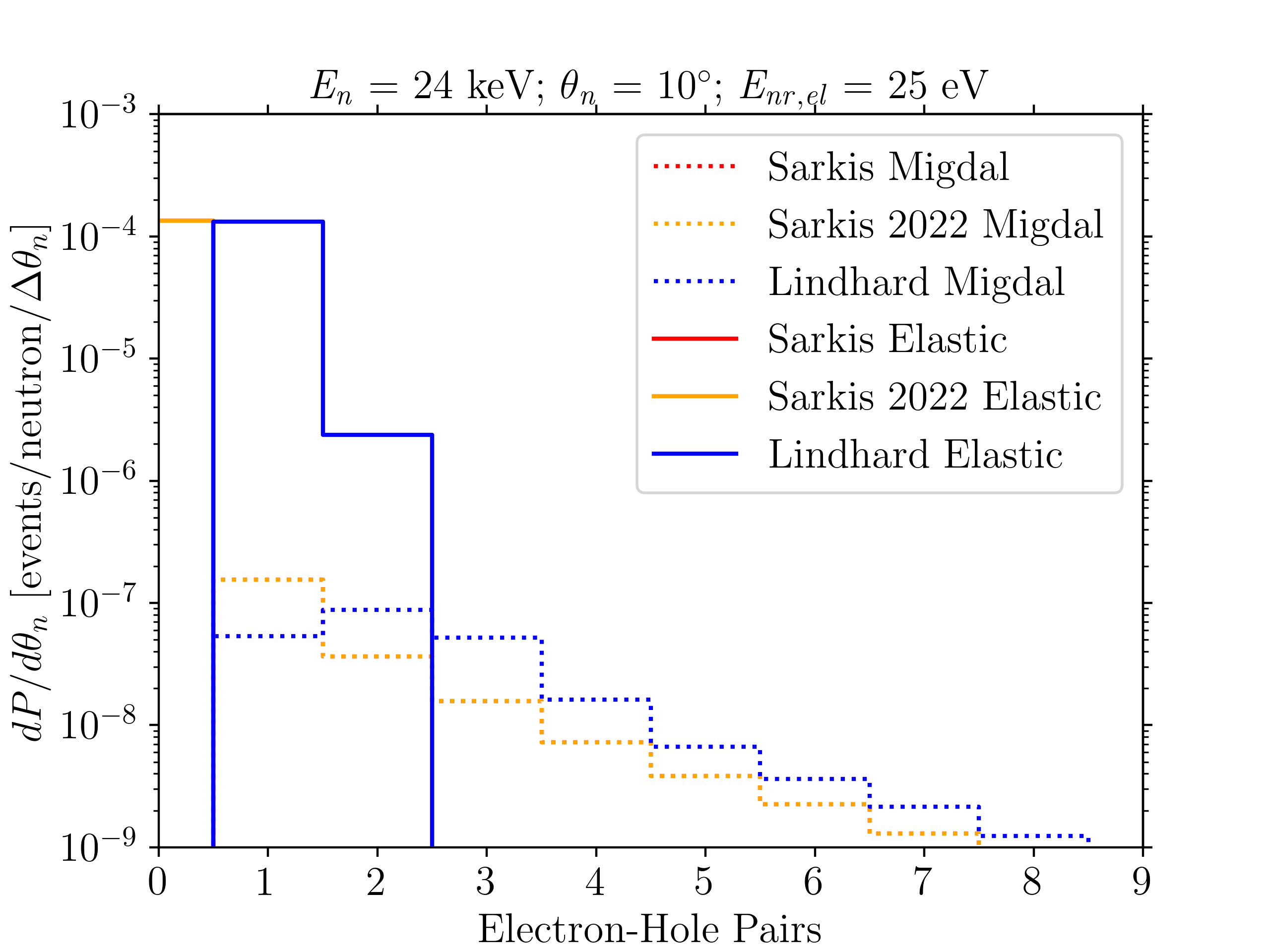}
\includegraphics[width=0.49\textwidth]{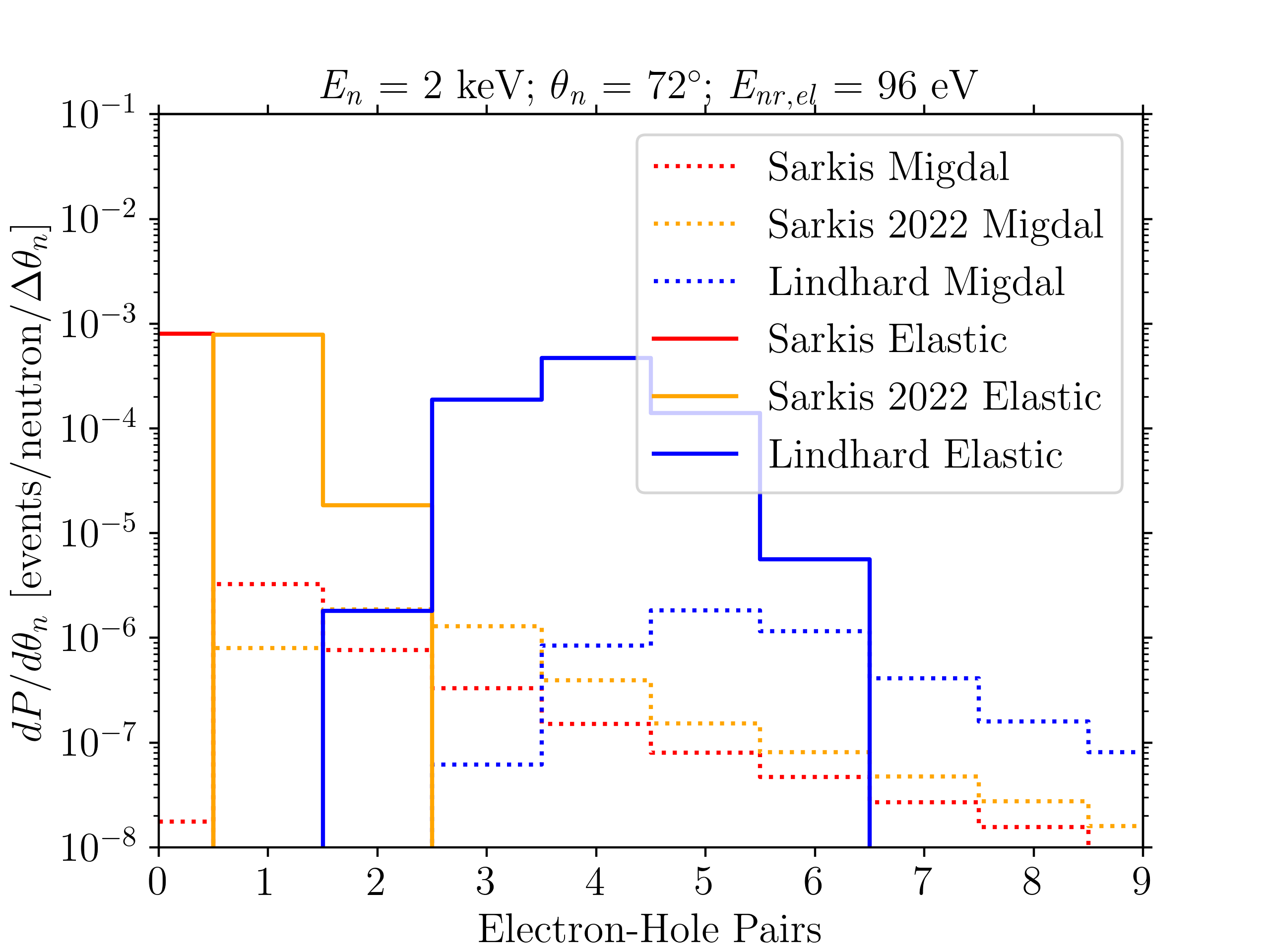}~
\includegraphics[width=0.49\textwidth]{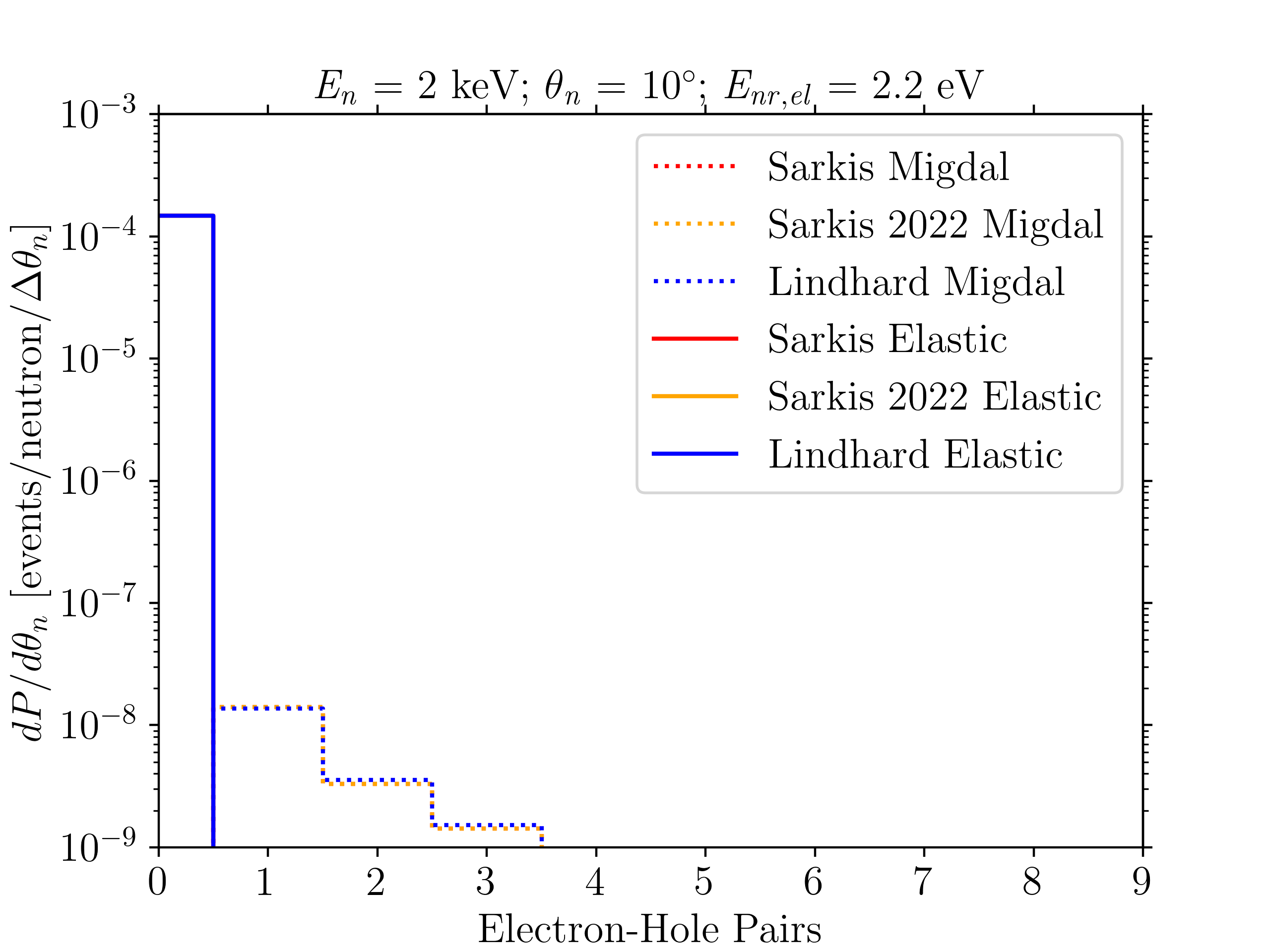}
\caption{
Each panel shows the differential probability spectra $dP_{n_e}/d\theta_n$ (in units of events/neutron/degree of angular coverage) per detectable charge quanta $n_e$ for elastic (solid) and Migdal (dashed) scattering expected for the Lindhard~\cite{osti_4701226} (blue), Sarkis~\cite{Sarkis:2020soy} (red), and Sarkis 2022~\cite{sarkis2022} (orange) ionization efficiency models and Ramanathan charge production model~\cite{Ramanathan:2020fwm} in a setup with an ideal neutron beam of energy $E_n$ incident on a 1~cm thick silicon detector with perfect detection efficiency. 
The left (right) column shows measured spectra for wide-(low-) angle neutron scattering at $\theta_n=$ 72 (10) degrees and the rows show, from top to bottom, the spectra for incident monoenergetic neutrons of 2.5~MeV (as from an unmoderated D-D generator), 24~keV (as from an iron filter), and 2~keV (as from a Sc filter). 
In all cases, we assume a perfect backing detector with full azimuthal coverage. 
Note that the high-energy, wide-angle case (top left) is outside of the soft-limit regime where our derivations are valid, giving the nonphysical amplification shown. On the other hand, the low-energy, low-angle case (bottom right) clearly demonstrates the scenario wherein the ME can be probed and measured independent of charge yield model using a single-electron sensitive detector.
}\label{fig:SpectrumAppendix}
\end{figure}

\begin{figure*}[!t]
\includegraphics[width=0.49\textwidth]{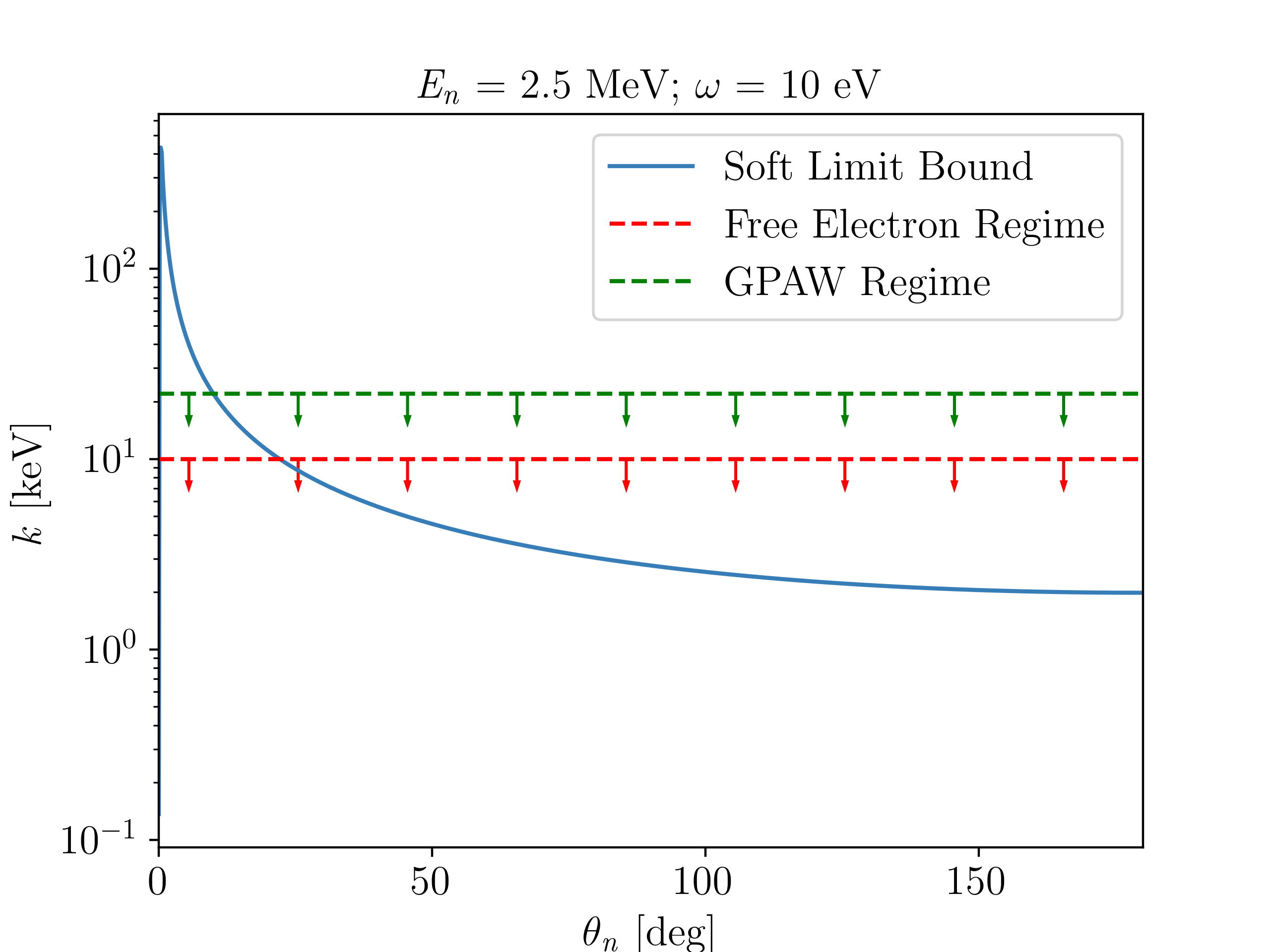}~
\includegraphics[width=0.49\textwidth]{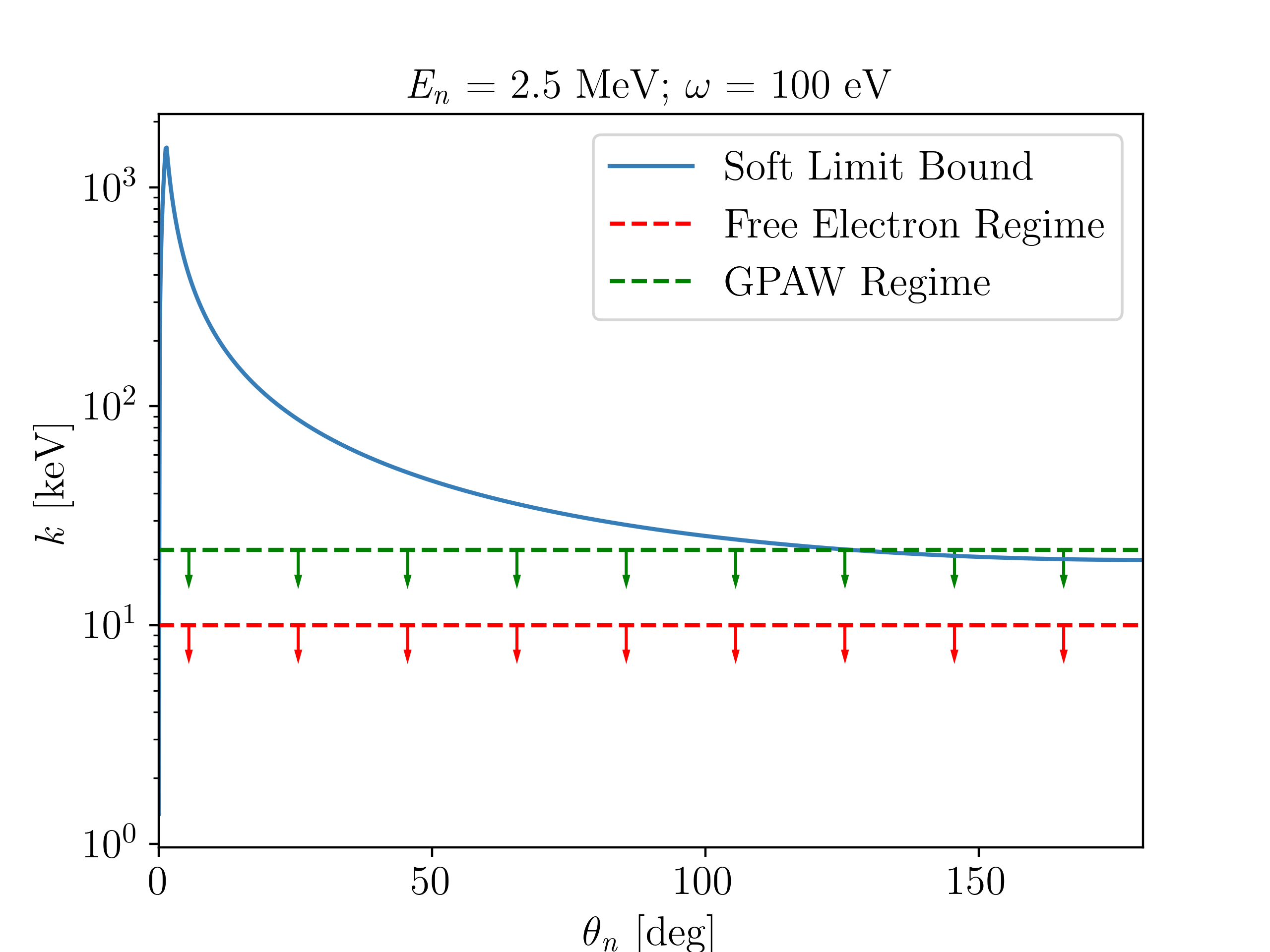}
\includegraphics[width=0.49\textwidth]{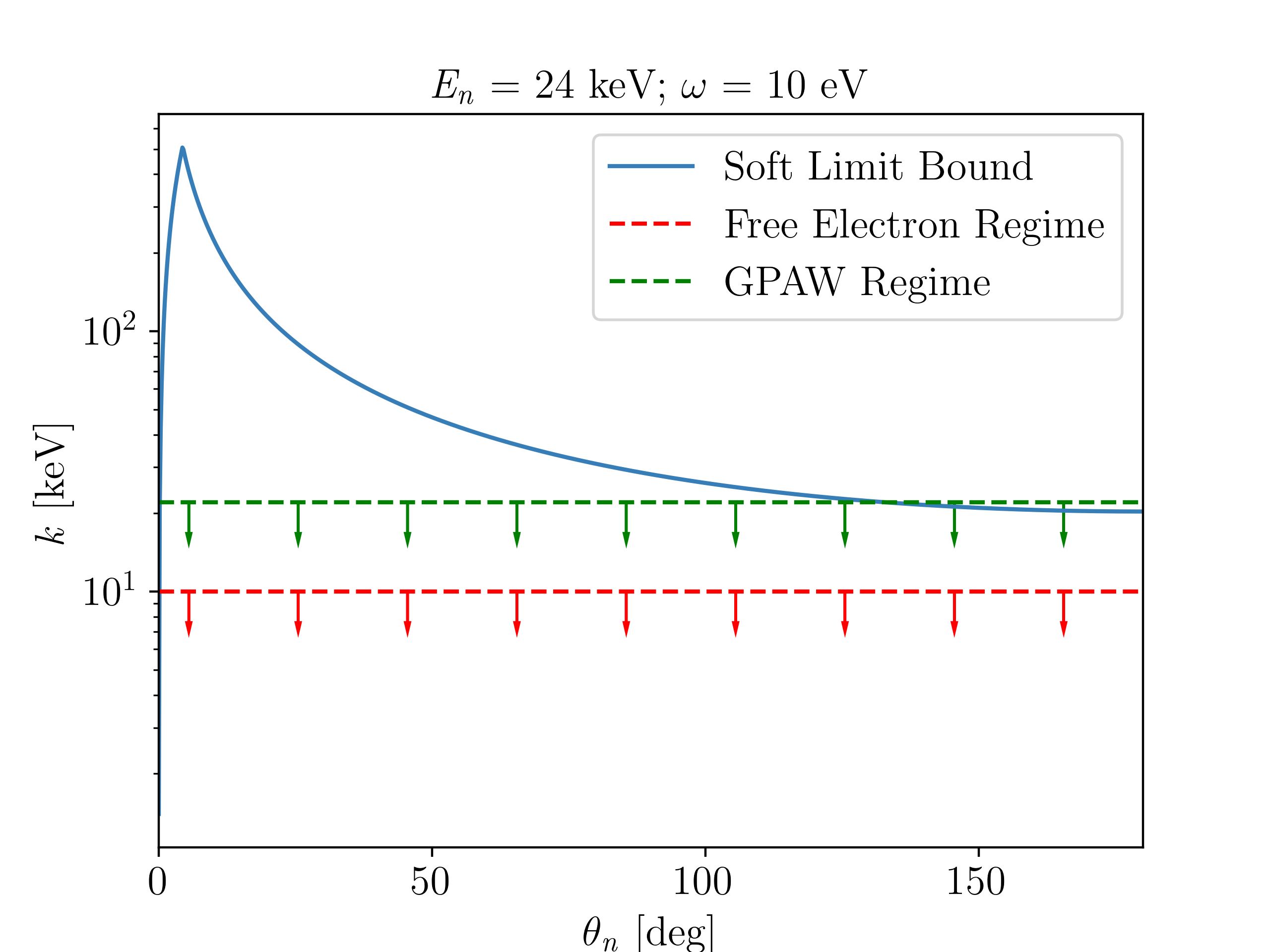}~
\includegraphics[width=0.49\textwidth]{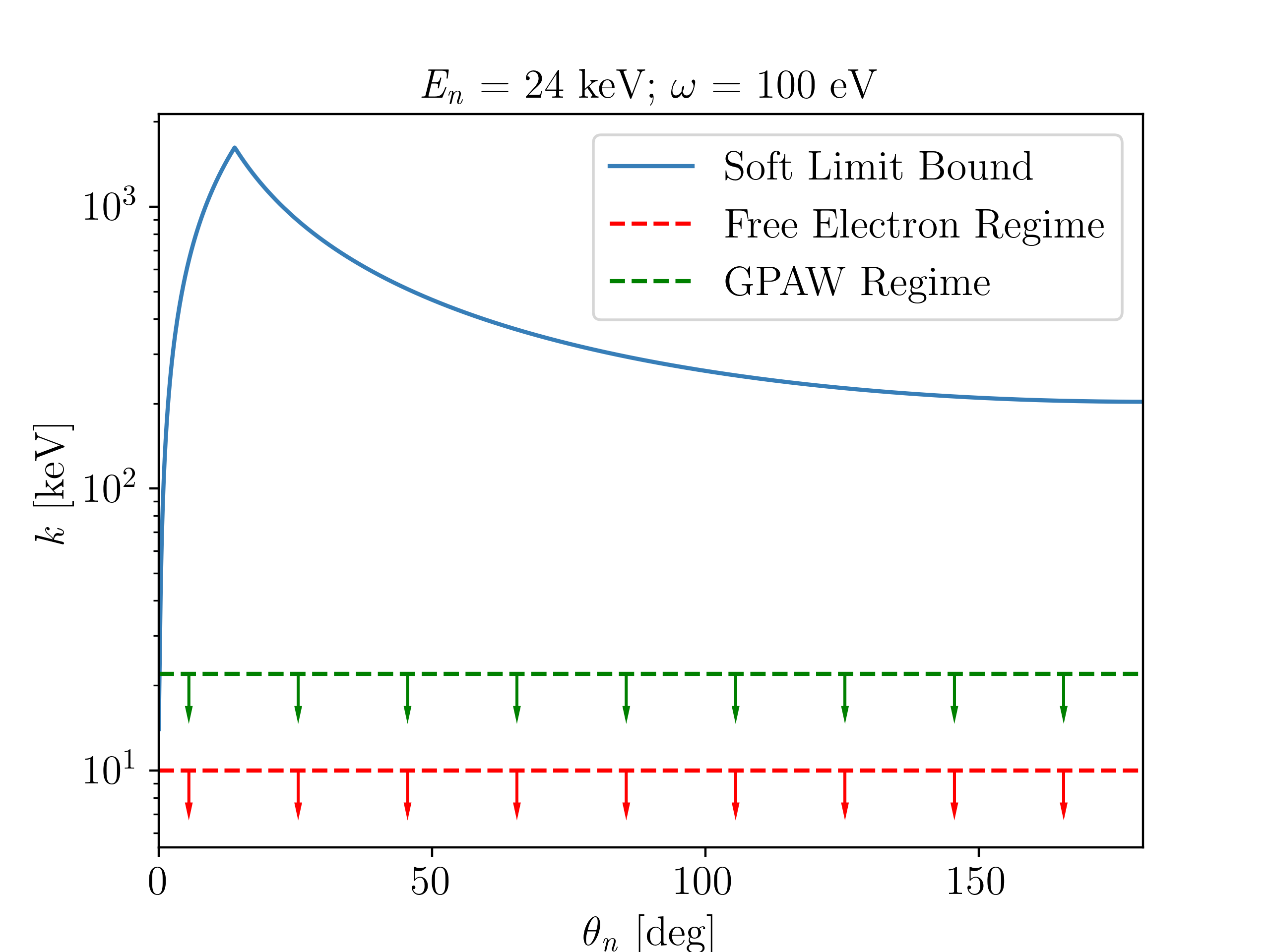}
\includegraphics[width=0.49\textwidth]{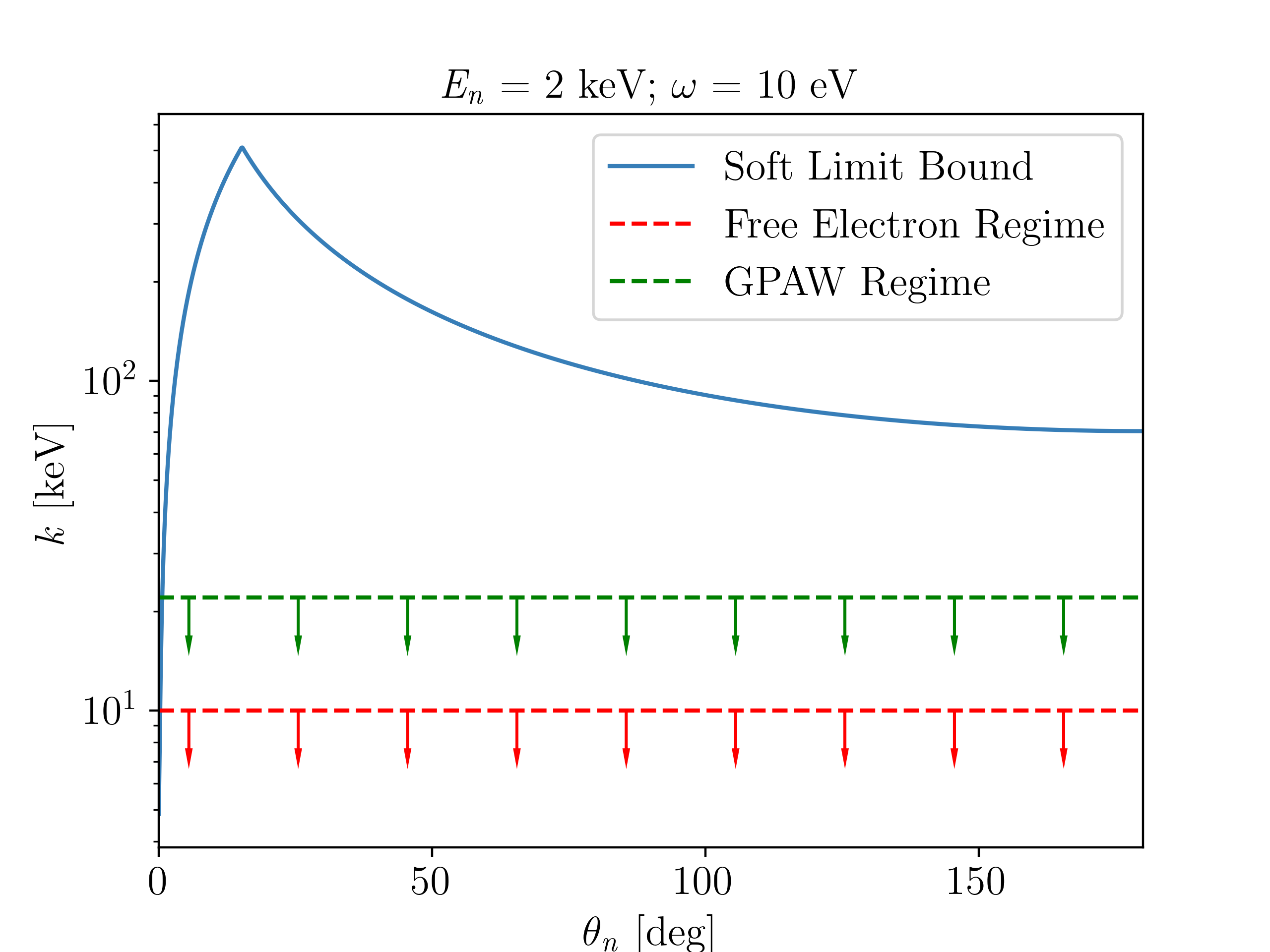}~
\includegraphics[width=0.49\textwidth]{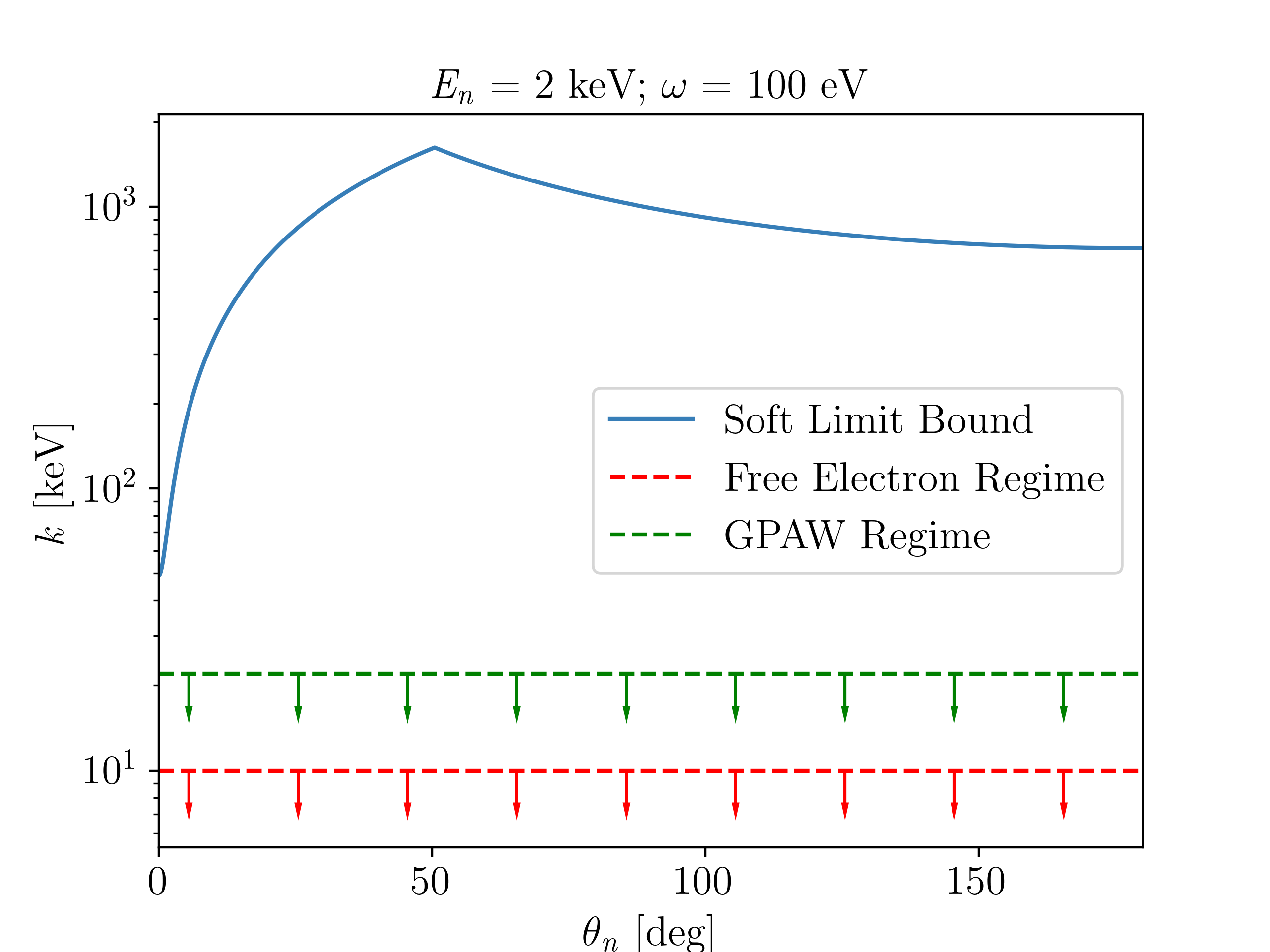}
\caption{
Regime of validity of the soft limit for the momentum transferred to the electronic system $k$ as a function neutron scattering angle $\theta_n$ for various incoming neutron energies $E_n$ and electronic energies $\omega$. The left (right) column shows momenta for low-(high-) energy transferred to the electronic system of $\omega = 10$ (100)~eV and the rows show, from top to bottom, the momenta for incident monoenergetic neutrons of 2.5~MeV (as from an unmoderated D-D generator), 24~keV (as from an iron filter), and 2~keV (as from a Sc filter). The red and green lines indicate the upper limit of the domain of validity for two models of the silicon ELF (see text for details). The blue line indicates the upper bound $k_{max}$ on for the soft limit approximation from Eq.~(\ref{eq:softcondition}). The soft limit calculations are expected to be valid so long as $k_{max}$ is larger than the domain of validity of the ELF; in particular, for high-energy neutrons (top row) and wide scattering angles at small $\omega$ (middle left), the soft limit is a poor approximation.}
\label{fig:softlim}
\end{figure*}

\section{Migdal Effect for Semiconductors Outside the Soft Limit}
\label{app:nosoft}
In this Appendix, we investigate how the kinematics of the Migdal effect change outside the soft limit. We start now with (A32) from Ref.~\cite{Knapen:2020aky}, suitably modified for neutron scattering:
\beq
\begin{aligned}
\frac{d\sigma}{d\omega} = C_n \int\frac{d^3 \vec{q}_N}{(2\pi)^3}\int\frac{d^3\vpf}{(2\pi)^3}\int\frac{d^3\vec{k}'}{(2\pi)^3}&(2\pi)^3 \delta^{3}(\vpi-\vpf-\vec{q}_N-\vec{k}') \delta\left( E_i-E_f-\omega-\tfrac{q_N^2}{2m_N}\right)\\
\times&Z_{\text{ion}}^2(\vec{k}')\, \frac{\text{Im}(-\epsilon^{-1}(\vec{k}', \omega))}{|\vec{k}'|^2}\left( \frac{1}{\omega-\frac{\vec{q}_N\cdot \vec{k}'}{m_N}}-\frac{1}{\omega}\right)^2\,.
\label{eq:notstart}
\end{aligned}
\eeq
As in Eq.~\eqref{eq:dmton}, 
we change the prefactor to $C_n$ appropriate for neutron scattering, abbreviate $\vec{k}' \equiv \vec{k} + \vec{K}$, and assume the free-ion approximation.
Here, we write $\vec{q}_N$ instead of $\vec{q}$ for the momentum transferred to the nucleus, because the electron system momentum $\vec{k}'$ appears in the momentum delta function. To facilitate comparison to the soft-limit derivation, we can first rewrite the expression in parentheses as
\beq
\left( \frac{1}{\omega-\frac{\vec{q}_N\cdot\vec{k}'}{m_N}}-\frac{1}{\omega}\right)^2 = \frac{1}{\omega^4m_N^2}\left( \frac{\vec{q}_N\cdot\vec{k}'}{1-\frac{\vec{q}_N\cdot\vec{k}'}{\omega m_N}}\right)^2\,.
\eeq

\subsection{Soft and low-momentum limits}

The two components of the soft limit correspond to dropping $\vec{k}'$ from two different parts of the integrand in Eq.~(\ref{eq:notstart}). Specifically, 
\begin{align}
\tag{SA}
k' \ll q_N & \implies \delta^{3}(\vpi-\vpf-\vec{q}_N-\vec{k}') \to \delta^{3}(\vpi-\vpf-\vec{q}_N)\,; \\
\tag{SB}
\vec{q}_N\cdot\vec{k}' \ll m_N \omega & \implies \left( \frac{\vec{q}_N\cdot\vec{k}'}{1-\frac{\vec{q}_N\cdot\vec{k}'}{\omega m_N}}\right)^2 \to (\vec{q}_N \cdot \vec{k}')^2\,.
\end{align}
Assuming only condition (SA) and isotropy of the target, we can perform identical manipulations to those in Appendix~\ref{app:softlimit}, and Eq.~(\ref{eq:notstart}) reduces to
\beq
\begin{aligned}
\frac{d\sigma}{d\omega d\cos\theta_n} = \frac{C_n}{m_n m_N \omega^4} \int& \frac{k'^2 dk' d \cos \theta_k}{(2\pi)^4}Z_{\text{ion}}^2(k')\mathcal{W}(k', \omega)\\
&\times\frac{p_f^2|\vpi-\vpf|^2 \cos^2 \theta_k}{\sqrt{m_n^2p_i^2\cos^2\theta_n+(m_N^2-m_n^2)p_i^2-2m_n(m_n+m_N)m_N\omega}}\left(\frac{1}{1-\frac{k'|\vpi-\vpf|\cos \theta_k}{m_N\omega}}\right)^2\,,
\end{aligned}
\label{eq:lowmomentum}
\eeq
where $|\vec{p_i} - \vec{p}_f| = \sqrt{p_i^2 + p_f^2 - 2 p_i p_f \cos \theta_n}$ and $p_f$ is given by $p_f^+$ in Eq.~(\ref{eq:pfpm}). We refer to the result obtained using only (SA) as the \emph{low-momentum limit}. At this point it is clear that unlike in the case of the soft limit, the angular dependence of the scattered neutron does not factorize from the electronic spectrum, even in the limit of an isotropic material, because of the presence of the term coupling $\cos \theta_k$ and $\omega$, which cannot be ignored without assumption~(SB).

Since $\vec{k}'$ is not observable and is, in fact, integrated over in the rate, a strict application of the soft limit effectively restricts the range of integration of $\vec{k}'$ for fixed momentum transfer $\vec{q}$ and electronic energy $\omega$. Since $|\vec{q} \cdot \vec{k}'| = \mathcal{O}(qk')$ unless $\vec{q}$ and $\vec{k}'$ are very nearly orthogonal (which, for an isotropic material, would suffer a $1/(4\pi)$ suppression in the angular part of the $\vec{k}'$ integral), the soft limit is roughly equivalent to
\begin{equation}
k' \ll {\rm min}\left( \sqrt{\frac{\omega^2 m_N}{2 E_r}}, \sqrt{2 m_N E_r}\right)\,,
\label{eq:softcondition}
\end{equation}
where we have replaced $q$ with $\sqrt{2 m_N E_r}$. To make contact with the kinematics discussed in the main text, we can write $E_r$ as a function of $E_n$, $\theta_n$, and $\omega$ using Eq.~(\ref{eq:ervstheta}). 
In Fig.~\ref{fig:softlim}, we explicitly plot the soft limit condition (\ref{eq:softcondition}) for the kinematics considered in Fig.~\ref{fig:SpectrumAppendix}. The two branches of $k_{\rm max}$ correspond to condition (SA) at small angles and (SB) at large angles, and the red and green horizontal lines correspond to the regimes of validity of two models for the ELF in silicon, an isotropic free-electron gas (see e.g. Ref~\cite{dressel2002electrodynamics}) and the \texttt{GPAW} model implemented in \texttt{DarkELF}, respectively. As noted in Ref.~\cite{Hochberg:2021pkt}, the free-electron gas model with Fermi velocity $v_F = 8.6 \times 10^{-3} c$  and plasmon frequency $\omega_p = 18.5 \ {\rm eV}$ is a reasonable approximation for the measured ELF in silicon for $k \lesssim 10 \ {\rm keV}$ and $5 \ {\rm eV} \lesssim \omega \lesssim 30 \ {\rm eV}$, capturing, in particular, the Fermi-broadened free-electron peak at $k \simeq \sqrt{2 m_e \omega}$. Deviations from the soft limit are expected when the blue curve drops below the red and green lines, which occurs either for very small scattering angles or for wider angles with sufficiently large $E_n$ and small $\omega$ (top row and middle left).

We now argue that the low-momentum limit matches the full result without (SA) very closely for our entire parameter space, which is convenient since Eq.~(\ref{eq:lowmomentum}) is easily amenable to numerical integration given a loss function $\mathcal{W}(k', \omega)$ . The relevance of (SB) but not (SA) for the soft limit can already be seen from Fig.~\ref{fig:softlim}, where (SA) is only violated for $\theta_n \ll 1^\circ$ for all choices of $E_n$ and $\omega$. To be somewhat more quantitative, we adopt the simple free-electron gas model of the ELF described above, which has a closed-form analytic expression~\cite{dressel2002electrodynamics}. However, it features an unphysical vanishing of the ELF at large $k$ where core electron wave functions should have nonvanishing support. A proper treatment of the ELF in silicon would include the effects of core electrons through, for example, ``all-electron reconstruction'' \cite{Griffin:2021znd}, which would eliminate the need to use the isolated atom formalism to compute the Migdal spectrum at large $\omega$. We leave this for future work. For the $k'$-dependent ion charge, we use an atomic form factor model,
\beq
Z_{\rm ion}(k') = \frac{Z_0 (\lambda_{\rm TF} k')^2}{1 + (\lambda_{\rm TF} k')^2}\,,
\eeq
where $\lambda_{\rm TF} = v_F/(\sqrt{3} \omega_p)$ is the Thomas-Fermi screening length and $Z_0 = 4$ is the charge of the silicon ion excluding the valence shell.

Fig.~\ref{fig:regimes} shows the effect on the angular spectrum $d\sigma/d\cos \theta_n$ of the extra term that would vanish in the full soft limit (SB). We have deliberately chosen kinematics that maximize the effect of (SB). The soft limit is expected to fail when the largest value of $k$ allowed by (SB) drops below the region where the ELF has large support. In the case of both the free-electron gas and \texttt{GPAW} ELFs, the ELF vanishes identically outside the regime of validity shown in Fig.~\ref{fig:softlim} above, and thus deviations from the soft limit are largest when the bound from (SB) is close to the limit of validity of the model. A more detailed calculation that accounts for inner electron shells in the ELF would not feature a hard cutoff in $k$ but rather something closer to a power-law falloff from momentum-space atomic orbitals, but the general phenomenon we illustrate here will still hold. 

The blue curve in Fig.~\ref{fig:regimes} shows the soft limit from Appendix~\ref{app:softlimit}, the green curve shows the low-momentum limit from Eq.~(\ref{eq:lowmomentum}), and the orange curve shows the full calculation of Eq.~(\ref{eq:notstart}), the details of which are given in Sec.~\ref{sec:nosoftfull} below. The kinematics are chosen to match Fig.~\ref{fig:softlim}, middle left and top right. There are order-1 deviations from the soft limit at wide scattering angles, but the low-momentum limit only differs from the full calculation at the percent level for any scattering angle. Furthermore, the soft limit underestimates the full result, because when (SB) does not hold, the nucleus propagator is closer to on shell.

\begin{figure}[t!]
    \centering
    \includegraphics[width=0.5\textwidth]{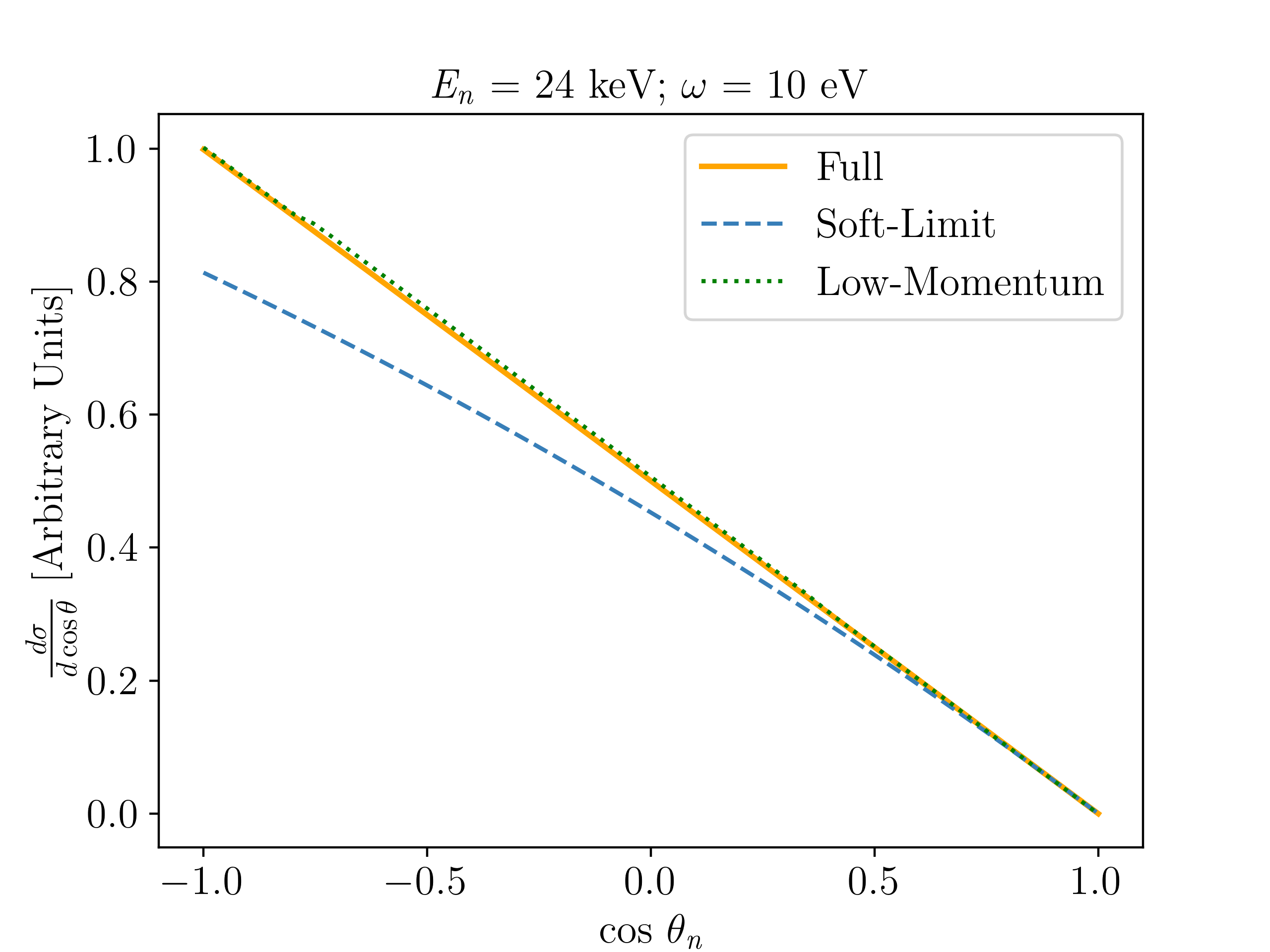}~
    \includegraphics[width=0.5\textwidth]{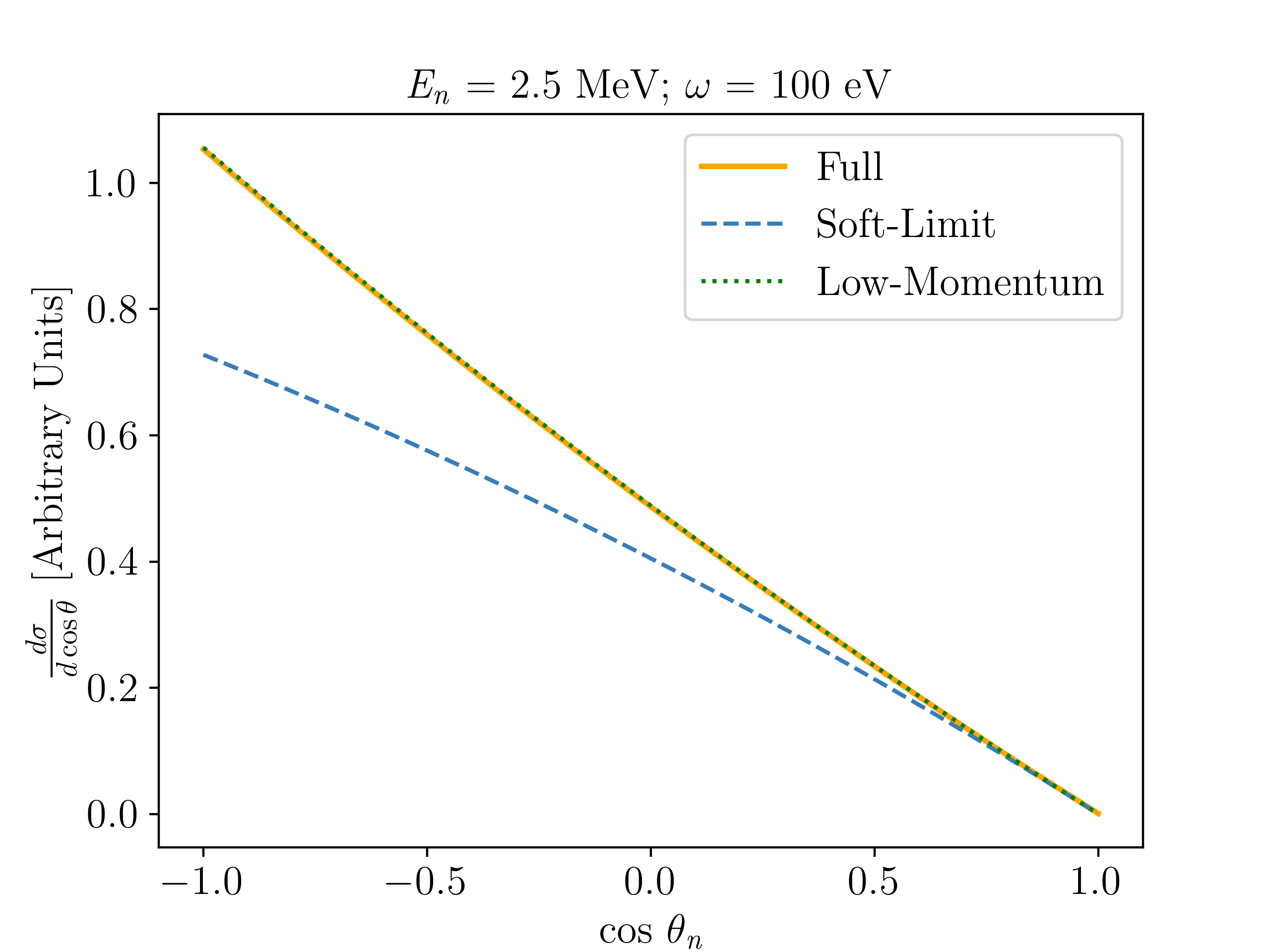}
    
    \caption{
    Comparison of the differential Migdal angular cross section in silicon for valence electrons in the soft limit and without assumptions (SA) and/or (SB) of the soft limit. The blue line is the full soft limit, the green line assumes only (SA) (low-momentum limit), and the orange line is the full result. The deviations from the soft limit are largest at wide scattering angles, and the low-momentum limit is nearly indistinguishable from the full calculation.
    The left (right) plot demonstrates this behavior for the case of medium-(high-) energy neutrons of $E_n = 24$~keV (2.5~MeV) as from the case of a iron filter (unmoderated D-D generator) for electronic energy transfers $\omega = 10$ (100)~eV.
    The comparison for cases in this Letter with large $\omega$ and small $E_n$ are not shown, as there is no significant deviation between the different calculations. 
    }
    \label{fig:regimes}
\end{figure}

\subsection{Full calculation outside the low-momentum limit}
\label{sec:nosoftfull}

For completeness, we now evaluate the full expression for the Migdal angular spectrum without assumption (SA). Starting from Eq.~(\ref{eq:notstart}), we can immediately perform the $\vec{q}_N$ integral with the momentum delta function by making the replacement $\vec{q}_N=\vpi-\vpf-\vec{k}'$. This leaves
\beq
\frac{d\sigma}{d\omega} = C_n \int\frac{d^3\vpf}{(2\pi)^3}\int\frac{d^3\vec{k}'}{(2\pi)^3}\delta\left( E_i-E_f-\omega-\tfrac{(\vpi-\vpf-\vec{k}^\prime)^2}{2m_N}\right) Z_{\text{ion}}^2(\vec{k}')\frac{\mathcal{W}(k', \omega)}{k'^2}\left( \frac{1}{\omega-\frac{(\vpi-\vpf-\vec{k}')\cdot\vec{k}'}{m_N}}-\frac{1}{\omega}\right)^2.
\eeq
To evaluate the remaining delta function, we use the nonrelativistic dispersion relation for the neutron, so that the delta function enforces
\beq
\frac{p_i^2-p_f^2}{2m_n} - \omega - \frac{(\vpi-\vpf)^2+k'^2-2(\vpi-\vpf)\cdot\vec{k}'}{2m_N} = 0, 
\eeq
or equivalently
\beq
-\left(\frac{1}{2m_n}+\frac{1}{2m_N}\right) p_f^2 + \frac{p_i\cos\theta_n-k'\cos\theta_{fk'}}{m_N}p_f + \left(\frac{1}{2m_n}-\frac{1}{2m_N}\right) p_i^2+\frac{p_ik'\cos\theta_{ik'}}{m_N}-\frac{k'^2}{2m_N}-\omega=0,\label{gy}
\eeq
where $\theta_{ik'}$ is the angle between $\vec{p}_i$ and $\vec{k}'$, and $\theta_{fk'}$ is the angle between $\vec{p}_f$ and $\vec{k}'$. Rearranging, we have
\beq
\frac{p_f^2}{2}-m_n\frac{p_i\cos\theta_n -k'\cos\theta_{fk'}}{m_n+m_N}p_f+\frac{m_n-m_N}{m_n+m_N}\frac{p_i^2}{2}+\frac{m_n}{m+m_N}\left( \frac{k'^2}{2}+m_N\omega-p_ik'\cos\theta_{ik'}\right)=0,
\eeq
which has the solution
\beq
p_f^\pm=m_n\frac{p_i\cos\theta_n-k'\cos\theta_{fk'}}{m_n+m_N}\pm\sqrt{\left( m_n\frac{p_i\cos\theta_n-k'\cos\theta_{fk'}}{m_n+m_N}\right)^2-\frac{(m_n-m_N)p_i^2+m_n(k'^2+2m_N\omega-p_ik'\cos\theta_{ik'})}{m_n+m_N}}. \label{soln}
\eeq
The solution $p_f^-$ becomes negative---and thus spurious---when
\beq
(m_N-m_n)p_i^2+m_n(p_ik'\cos\theta_{ik'}-k'^2-2m_N\omega)>0,
\eeq
which is always the case except very close to threshold, $\omega\approx E_i$. To proceed in full generality, though, we will keep $p_f^-$.

As in the soft limit derivation, the square root gives us the constraint
\beq
\frac{m_n^2}{m_n+m_N}(p_i\cos\theta_n -k'\cos\theta_{fk'})^2>(m_n-m_N)p_i^2+m_n(k'^2+2m_N\omega-2p_ik'\cos\theta_{ik'}),
\eeq
which now implies a restriction on the integration range of $k'$:
\beq
\left(1-\tfrac{m_n}{m_n+m_N}\cos^2\theta_{fk'}\right) k'^2 + 2p_i\left(\tfrac{m_n}{m_n+m_N}\cos\theta_n\cos\theta_{fk'}-\cos\theta_{ik'}\right) k'+\left( 1-\tfrac{m_n}{m_n+m_N}\cos^2\theta_n -\frac{m_N}{m_n}\right) p_i^2 +2m_N\omega <0 \label{inequal}
\eeq
The solution to this inequality gives lower and upper bounds on $k'$. As long as $p_i^2 \gg \omega$ and $m_N > m_n$ (which is always true for the scenarios we consider), the lower bound on $k'$ is negative and therefore spurious, and the upper bound is much larger than the domain of validity of the valence-shell ELF model, $k_{\rm max} \simeq 30 \ {\rm keV}$. So in practice, the energy-conserving delta function does not restrict the 3-body kinematics.

Using the delta function to perform the $p_f$ integral, we identify the Jacobian via 
\beq
\delta(g(x))=\sum_i\frac{\delta(x-x_i)}{|g'(x_i)|},
\eeq
where the $x_i$ are the roots of $g(x)$ and the derivative is with respect to the argument $x_i$. For the delta function in question, $g(x)$ is given by the left-hand side of Eq.~(\ref{gy}) and the $x_i$ are $p_f^+$ and $p_f^-$, so we get
\beq
\begin{aligned}
|g^\prime(p_f)|&=\frac{1}{m_N}\left|p_i\cos\theta_n -k'\cos\theta_{fk'}-\frac{m_n+m_N}{m_n}p_f\right|\\
&=\frac{1}{m_n m_N}\sqrt{m_n^2(p_i\cos\theta_n -k'\cos\theta_{fk'})^2+(m_N^2-m_n^2)p_i^2-m_n(m_n+m_N)(k'^2+2m_N\omega-p_ik'\cos\theta_{ik'})}.
\end{aligned}
\eeq
Combining everything together and simplifying, we get
\beq
\frac{d\sigma}{d\cos\theta_n d\omega} = \frac{C_n m_N}{\omega^2}\int\frac{d^3\vec{k}'}{(2\pi)^5} Z_{\text{ion}}^2(k')\frac{\mathcal{W}(k', \omega)}{k'^2}
\left( \sum_{+, -} \frac{(p_f^\pm) ^2\left(\frac{p_i\cos\theta_{ik'}-p_f^\pm \cos\theta_{fk'}-k'}{\frac{m_N\omega}{k'}-p_i\cos\theta_{ik'}+p_f^\pm \cos\theta_{fk'}+k'}\right)^2}{\left|p_i\cos\theta_n-k'\cos\theta_{fk'}-\frac{m_n+m_N}{m_n}p_f^\pm\right|}
\right).
\label{eq:not}
\eeq
where the sum is over the two terms containing $p_f^+$ and $p_f^-$. To perform the remaining angular integrals, we note that the remaining symmetry axis is along $\vec{p}_i-\vec{p}_f$, which varies with $p_f$ through its dependence on $k$ and $\theta_{ik}$. The length $d$ of $\vec{p}_i-\vec{p}_f$ can be determined using the law of cosines:
\beq
(d^\pm)^2=p_i^2+(p_f^\pm)^2-2p_ip_f^\pm \cos\theta_n
\eeq
as can the angle $\theta_{fd}$ between $\vec{p}_f$ and the symmetry axis:
\beq
p_i^2=(p_f^\pm)^2+d^2-p_f^\pm d\cos\theta_{fd}^\pm.
\eeq
Using these relations, we can define the polar angle with respect to the symmetry axis
\beq
\theta^\pm=\theta_{ik'}+\pi-\theta_n-\theta_{fd}^\pm(\theta_{ik'}).
\eeq
This is an implicit equation for $\theta_{ik'}$ which must be solved to make the variable substitution necessary to integrate over $\theta^\pm$. Then the remaining $\phi$ integral is trivially integrated about the symmetry axis and results in a factor of $2\pi$, and we can define the remaining $\theta$ integration variable to be $\xi=\theta^+=\theta^-$. Finally, noting that 
\beq
p_i \cos\theta_{ik'}-p_f^\pm\cos\theta_{fk'}=\frac{\vec{p}_i\cdot \vec{k}'-\vec{p}_f^\pm\cdot \vec{k}'}{k'}=|\vec{p}_i-\vec{p}_f^\pm|\cos\theta^{\pm},
\eeq
Eq.~(C16) reduces to 
\beq
\begin{aligned}
\frac{d\sigma}{d\omega d\cos\theta_{if}} = C_n \int& dk'd\xi\frac{\sin\xi}{(2\pi)^4} Z_{\text{ion}}^2(k')\mathcal{W}(k', \omega)\frac{k'^2}{mm_N\omega^4}\sum_{+, -}\frac{(p_f^\pm(\xi))^2(|\vpi-\vpf^\pm(\xi)|\cos\xi-k')^2}{\left(1-\frac{\vec{k}'\cdot(\vec{p}_i-\vec{p}_f^\pm(\xi)-\vec{k}')}{m_N\omega}\right)^2}\\
&\times\frac{1}{\sqrt{m_n^2(p_i\cos\theta_n-k'\cos\theta_{fk'}(\xi))^2+(m_N^2-m_n^2)p_i^2-m_n(m_n+m_N)(k'^2+2m_N\omega-p_ik'\cos\theta_{ik'}(\xi))}}.
\end{aligned}\label{nots}
\eeq
From this expression, we confirm that the angular spectrum does not factorize outside the soft limit.

\end{appendix}

\end{widetext}